\documentclass[sn-mathphys]{sn-jnl} 

\usepackage{amssymb,hyperref,amsthm,amsmath}
\usepackage{epstopdf}
\usepackage{url}
\usepackage[normalem]{ulem}
\usepackage{etoolbox}
\usepackage{cprotect}
\usepackage{graphicx}
\usepackage{xcolor}

\theoremstyle{plain}
\newtheorem{theorem}{Theorem}[section]
\newtheorem{proposition}[theorem]{Proposition}
\newtheorem{definition}[theorem]{Definition}
\newtheorem{example}[theorem]{Example}
\newtheorem{remark}[theorem]{Remark}

\begin{document}

\title{A topological perspective on weather regimes}

\author*[1]{Kristian Strommen}
\email{kristian.strommen@physics.ox.ac.uk}

\author[1]{Matthew Chantry}\email{matthew.chantry@physics.ox.ac.uk}

\author[1]{Joshua Dorrington}\email{joshua.dorrington@physics.ox.ac.uk}

\author[2]{Nina Otter}
\email{n.otter@qmul.ac.uk}

\affil[1]{University of Oxford, Department of Physics, Parks Rd, Oxford OX1 3PJ, United Kingdom}
\affil[2]{UCLA, Department of Mathematics, Los Angeles, CA, USA 90095}

\date{\today}

\abstract{It has long been suggested that the mid-latitude atmospheric circulation possesses what has come to be known as `weather regimes', loosely categorised as regions of phase space with above-average density and/or extended persistence. Their existence and behaviour has been extensively studied in meteorology and climate science, due to their potential for drastically simplifying the complex and chaotic mid-latitude dynamics. Several well-known, simple non-linear dynamical systems have been used as toy-models of the atmosphere in order to understand and exemplify such regime behaviour. Nevertheless, no agreed-upon and clear-cut definition of a `regime' exists in the literature. We argue here for an approach which equates the existence of regimes in a dynamical system with the existence of non-trivial topological structure of the system's attractor. We show using persistent homology, an algorithmic tool in topological data analysis, that this approach is computationally tractable, practically informative, and identifies the relevant regime structure across a range of examples.}

\maketitle

\section{Introduction}

The discovery of the `butterfly effect' \cite{Lo63a} effectively ended the idea that weather forecasting can be understood purely as the problem of integrating a deterministic system forward in time. Instead, the problem of accurate weather forecasting becomes one of determining, from a given initial state, the likely trajectories of the atmosphere on its underlying attractor \cite{Slingo2011}. Similarly, the problem of producing reliable climate projections can be understood as determining how, and to what extent, the likelihood of traversing different trajectories changes in the presence of an external forcing \cite{Corti1999, Palmer1999, Woollings2010a}. As such, it becomes natural to ask whether the climate attractor exhibits significant deviations from Gaussianity, since such deviations, even locally, may strongly constrain the available trajectories. In other words, understanding the `shape' of the attractor becomes a problem of great practical importance.

\begin{figure}[h!]
\centering
\includegraphics[scale=0.55]{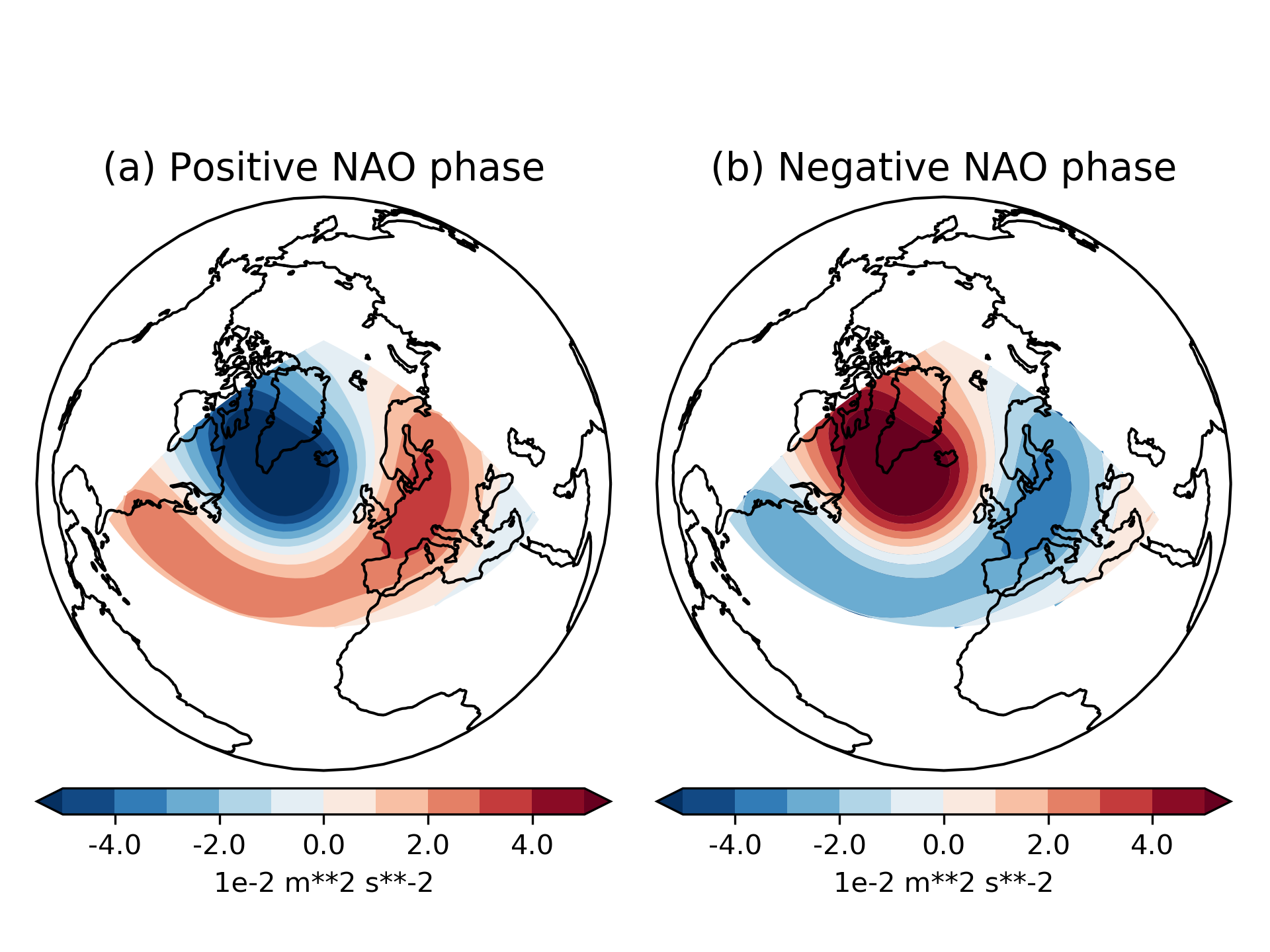}
\caption{The positive (a) and negative (b) phases of the North Atlantic Oscillation (NAO), the dominant pattern of variability in the North Atlantic wintertime circulation. The pattern is a dipole of atmospheric pressure anomalies, measured here as the first empirical orthogonal function of geopotential height at 500hPa. The two phases can be viewed as one of several possible ways to decompose the North Atlantic circulation into distinct regimes.}
\label{fig:nao_patterns}
\end{figure}

The study of local non-Gaussianity in the atmosphere has classically been done under the guise of so-called weather (or circulation) regimes \cite{Vautard1990, Michelangeli1995, Corti1999,Lorenz2006}. The basic idea is to determine a small number of dynamically relevant large-scale flow patterns (the regimes) that dominate the low-frequency variability and transition from one to another in an approximately Markovian manner \cite{baur1951extended}. We note that the use of the word `regime' in this paper should not be confused with usages associated with transitions induced by changes in the system's parameters, i.e., bifurcations. This concept of atmospheric weather regimes has been most famously applied to study the circulation of the wintertime North Atlantic sector. Here, the dominant pattern of variability is a dipole pattern of pressure anomalies referred to as the North Atlantic Oscillation (NAO) \cite{Hurrell2003}, shown in Figure \ref{fig:nao_patterns}. Knowing whether the NAO is in its positive or negative phase gives a good first-order approximation of winter weather in Europe and eastern North America, and is one of several possible regime views of the North Atlantic circulation. Strong NAO events are frequently linked to extreme surface weather and increased predictability, such as the European winter of 2019/2020, the warmest on record to date \cite{Hardiman2020}. Simple dynamical systems such as the Lorenz `63  system \cite{Lo63a}, the Charney-deVore \cite{Charney1979} and Lorenz `96 systems \cite{Lorenz1996} have frequently been used to illustrate or study the use of such atmospheric weather regimes \cite{Palmer1999, Charney1979, Christensen2015}.

However, despite being studied since the 1950's \cite{baur1951extended}, no clear-cut and generally accepted definition of a regime exists. Most definitions found in existing studies, often stated only implicitly, are based either on density considerations, where a regime corresponds to a region of above-average density in phase space (`clusters') \cite{Stephenson04, Vautard1990, Straus2010}, or some temporal persistence criteria, whereby a regime is a dynamical phenomenon (e.g., a blocking anticyclone) with a clear lifecycle and a lifespan exceeding some prescribed threshold \cite{Mo1987, Lorenz2006, Franzke2008, Falkena2020}; some studies also combine the two \cite{Falkena2020}. When applied to the North Atlantic circulation, these approaches typically produce anywhere between 2 and 6 regimes which are not, in general, easily comparable with each other. The resulting ambiguity has led some to question whether meaningful weather regimes really exist at all \cite{Stephenson04, Christiansen2007, Fereday2017}.

Besides the lack of agreement in the definition of regimes, existing approaches suffer from two key problems. Firstly, the algorithms involved often require essentially ad-hoc choices up front, such as the choice of number of clusters in $K$-means clustering algorithms, or temporal persistence thresholds, which directly influence the output regimes. This adds an additional layer of complexity to analysis, as it can be hard to motivate the choice of one parameter over another \cite{Christiansen2007, Dorrington2020}. Secondly, and more seriously, as we will show, any simplistic definition based on density and temporal persistence invariably fails to account for one or more classical regime systems in the literature. While a more technical definition based on exact solutions (e.g., fixed-points and periodic orbits) of the flow can work and even be computationally tractable for relatively low-dimensional systems \cite{gibson2008visualizing, ding2016estimating}, they suffer from the `curse of dimensionality', and are generally limited to simple systems. In a state-of-the-art application of these concepts to atmospheric modelling, \cite{lucarini2020new} identified unstable periodic orbits (UPOs) corresponding to zonal and blocking events in a low resolution quasi-geostrophic model. However, the dimensionality of this model still sits well below that of weather and climate models, let alone the physical system itself. Inherently, UPOs and their stability are model features, limited by the accuracy of their associated models and such analysis cannot be directly applied to observational data.

One perspective, in response to these difficulties, is that a regime is simply a deliberately vague word useful for capturing a variety of different ways to simplify complex dynamics, where the specifics are guided by context. The emergence of the field of topological data analysis \cite{carlsson08} offers another perspective, by bringing the attention away from specific dynamical properties, such as density and temporal persistence, and back to a more general consideration of the `shape', i.e., the \emph{topology}, of dynamical systems. We will argue that the only clearly unifying feature shared between a number of classic examples of regime systems (Lorenz `63, Lorenz `96, Charney-deVore and the North Atlantic jet) is their non-trivial topology, and that the particular topological structure associated with these systems captures well their most familiar features. This suggests that regimes might be understood as the results of varied attempts to capture the non-trivial topology of the underlying attractor.

In order to compute such topological structure \emph{directly}, we make use of persistent homology \cite{otter2017roadmap}, a technique in topological data analysis that gives a principled way of computing homology groups, which encode topological invariants (e.g., Betti numbers), for point-cloud data sets. Given a point cloud, one associates to it a filtration of spaces, for instance by thickening each point with a ball of radius $r$, and one then computes the homology for increasing values of $r$. Persistent homology then, gives a summary, called a barcode, which encodes how topological invariants, such as components or holes, evolve across the filtration. Because many iconic topological features of dynamical systems, such as the two holes in the Lorenz `63 system (cf. Figure \ref{fig:lorenz_holes_evolution}), can effectively vanish in the limit of infinitely many points, it turns out to be crucial to augment the standard filtration by taking into account density. To do this, we compute, for any dynamical system, a \emph{bifiltration} of topological invariants, and we then use it to study  non-trivial topological features of the attractor, such as the number of connected components (i.e., well-separated regions of points), and holes (e.g., as in Lorenz `63). This bifiltration thereby gives a way of measuring the `regime structure' in a robust and computationally tractable way. In particular, non-trivial topological invariants can be viewed as indicating the existence of regimes.

Let us make some remarks on technical benefits. Firstly, our method does not require an up-front choice of parameters that directly influences the regime structure found, unlike for instance K-means clustering. Secondly, homological computations do not essentially depend on the dimensionality of the data set, meaning this method does not suffer from the usual `curse of dimensionality'. Finally, persistent homology is model-independent, in that it does not require any prior knowledge of the underlying equations defining the data set, unlike, for instance, UPOs. This method therefore provides, in principle, a way to search for regime structure in arbitrary high-dimensional data sets with no a priori knowledge about whether they have regimes or not.

The potential of persistent homology as a tool for analysing dynamical systems was first suggested, among others, in \cite{Maletic2016}, in which it was demonstrated that persistent homology can locate the holes of the Lorenz `63 system: see also \cite{Ghil2020}. Of particular relevance is the recent work of \cite{Gokhan2020}, which uses persistent homology and UPOs in order to obtain a simplified representation of chaotic dynamical systems, an approach similar in spirit to our paper. A recent application of persistent homology to the real atmosphere is \cite{Muszynski2019}, which studies `atmospheric rivers' using a combination of homology and machine learning.  For a  different application of topological ideas to ocean modeling, we can also recommend \cite{Stanley2019}. There are several additional lines of work, that have used  methods from topology to study dynamical systems, such as \cite{KM16}, and \cite{KLT16}, to cite a few.

Finally, a cautionary note on language. The use of the word `persistent' in `persistent homology' comes from the way in which topological features that persist for a certain number of filtration values are considered to give meaningful information. In particular, there is no obvious relationship with the temporal persistence of regime states. To avoid ambiguity, in this manuscript the word `persistence' will always refer to topological persistence, while when referring to temporal persistence of regime states, we will make use of the qualifier `temporal'. 

The paper is structured as follows. In Section \ref{sec:data} we provide details of the dynamical systems used, including the observational atmospheric data. In Section \ref{sec:background}, we provide an informal introduction to persistent homology, and we motivate the need for a bifiltration. The formal, mathematical definitions of the concepts underlying persistent homology are included in Appendix \ref{SS:definition of barcode}; readers willing to treat these formalities as a `black box' should not find their understanding of the paper otherwise compromised. In Section \ref{sec:methodology} we detail the algorithmic procedure used to compute topological metrics. The results of applying this methodology to our suite of data sets is shown and discussed in Sections \ref{sec:results} and \ref{sec:discussion}, with conclusions and future directions in Section \ref{sec:conclusions}.

\section{Data}
\label{sec:data}

\subsection{Lorenz `63}

The Lorenz `63 system, first introduced and studied in \cite{Lo63a}, is a chaotic dynamical system in three variables $x,y,z$ defined by the equations
\begin{eqnarray}
        \dot{x} &=& \sigma (y-x), \nonumber \\
        \dot{y} &=& x(\rho -z)-y, \nonumber \\
        \dot{z} &=& xy - \beta z, \nonumber
\end{eqnarray}
and represents a highly simplified model of Rayleigh-B{\'e}rnard convection. It was also re-derived as a toy model of the NAO \cite{Molteni2019}. The attractor famously resembles a butterfly, usually viewed as having two regimes corresponding to the two `wings'; its regime behaviour has been extensively studied \cite{Palmer1994, Yadav2005}. Here we use the standard choice of constantzs $\sigma, \beta$ and $\rho$, namely $\sigma=10, \beta=8/3, \rho=28$. We generate a timeseries of 20000 points  by integrating the equations with a forward Euler scheme at a timestep $dt=5\cdot 10^{-5}$.

\subsection{Charney--DeVore}
\label{sec:cdv}

The Charney-deVore (CdV) model, first derived in \cite{Charney1979}, provided one of the first examples of multiple invariant measures in an atmospheric model, and can be thought of as a crude model of large-scale midlatitude blocking dynamics. It is based on a severe spectral truncation of the barotropic vorticity equation in a $\beta$-plane channel, as in Equation (\ref{eq:master}), where $\Psi$ is a streamfunction, $\gamma h$ is an orographic profile and $\Psi^*$ is an external forcing.

\begin{equation}
\frac{\partial }{\partial t}\nabla^{2}\Psi=-J(\Psi, \nabla^{2}\Psi +\gamma h) -\beta\frac{\partial\Psi}{\partial x}-C(\Psi-\Psi^{*})
\label{eq:master}
\end{equation}

While in \cite{Charney1979} the main focus is on a three-mode truncation of the system, where a marginally less severe truncation keeping three zonal and two meridional modes is applied, the p.d.e.\ reduces to the six-equation o.d.e.\ system shown in Equation \ref{eq:xx}, containing quadratic non-linearities and linear coriolis, orographic, and relaxation terms.

\begin{eqnarray}\label{eq:xx}
\dot{x_{1}} &=& \tilde{\gamma_{1}}x_{3} -C(x_{1} -x_{1}^*) \nonumber \\
\dot{x_{2}} &=& \beta_{1} x_{3} -\alpha_{1}x_{1}x_{3} -\delta_{1}x_{4}x_{6} -C(x_{2} -x_{2}^*) \nonumber \\
\dot{x_{3}} &=& -\beta_{1} x_{2} -\gamma_{1}x_{1} +\alpha_{1}x_{1}x_{2} +\delta_{1}x_{4}x_{5} -C(x_{3} -x_{3}^*) \\
\dot{x_{4}} &=& \tilde{\gamma_{2}}x_{6} +\epsilon\cdot(x_{2}x_{6} - x_{3}x_{5}) -C(x_{4} -x_{4}^*) \nonumber \\
\dot{x_{5}} &=& \beta_{2} x_{6} -\alpha_{2}x_{1}x_{6} -\delta_{2}x_{3}x_{4} -C(x_{5} -x_{5}^*) \nonumber \\
\dot{x_{6 }} &=& -\beta_{2} x_{5} -\gamma_{2}x_{4} +\alpha_{2}x_{1}x_{5} +\delta_{2}x_{2}x_{4} -C(x_{6} -x_{6}^*) \nonumber
\end{eqnarray}

A parameter set where this model produces chaotic dynamics was found in \cite{Crommelin2004}, and we use those same parameters here (see ibid for a full discussion of the constant values and meaning of each term in equation \ref{eq:xx}). An interactive simulation showing the evolution of this system can be found at \url{joshdorrington.github.io/cdv_simulator/}.

This system has been introduced as it exhibits multimodality (i.e., regimes) in a model which is significantly more complex and more physically interpretable than the Lorenz `63 system, and which also has particularly challenging phase-space structure. The regime dynamics are of Pomeau-Maneville type \cite{Pomeau1980} in that they consist of long-lived quasi-stationary periods in the vicinity of a weakly unstable fixed point, punctuated by a `bursting' behaviour and a transition to chaotic flow. These chaotic transients shadow unstable homoclinic orbits radiating from the fixed point, and so lend  considerable structure to the model attractor, with a series of strongly preferred looping trajectories. A timeseries of 20000 points was generated by integrating the equations with a forward Euler scheme at a timestep $dt=2\cdot 10^{-4}$. In order to visualise the data in three dimensions, a truncation of the six dimensional space is required. Because around 98\% of the variance is explained by the first three empirical orthogonal functions (EOFs), we use these to define a truncated space. Homological computations were found to be essentially unchanged when using the truncated space or all six dimensions, so the truncated space is used in all computations.

\subsection{Lorenz `96}

The Lorenz `96 model was introduced in \cite{Lorenz1996} as an idealized, chaotic model of the atmosphere which is of greater complexity than the Lorenz `63 system \cite{Karimi2010}. It is defined in our case by coupling eight variables $X_k, k=1, \ldots, 8$, representing large-scale variability with 32 variables $Y_j, j=1,\ldots 32$, representing small-scale variability, using the following equations:

\begin{eqnarray}\label{l96eqs}
    \dot{X_k} &=& -X_{k-1}(X_{k-2} - X_{k+1}) - X_k + F 
                     - \frac{hc}{b} \sum_{j=J(k-1)+1}^{kJ} Y_j; \, k=1, \ldots, K, \\
    \dot{Y_j} &=& -cbY_{j+1}(Y_{j+2} - Y_{j-1}) - cY_j 
               + \frac{hc}{b}X_{int[(j-1)/J]+1}; \, j=1,\ldots,JK. 
\end{eqnarray}
Cyclic boundary conditions are then imposed: $X_{k+K} = X_k, Y_{j+jK}=Y_{j}$. The parameters  are chosen as in \cite{Christensen2015}, which also discusses the meaning of the different constants.

Due to the interpretation of the equations in terms of large-scale modes coupled to small-scale modes, Lorenz `96 has been utilised in several studies looking at different ways to parameterise unresolved sub-grid scale variability in forecast systems \cite{Wilks2005, Christensen2015, Vissio2018, Gagne2020}. Its regime structure has been considered in, e.g., \cite{Lorenz2006} and \cite{Christensen2015}. The analysis of ibid also makes it clear that the key regime variability is concentrated in the first four EOFs. Computations are therefore always done using the subspace spanned by these.

\subsection{Observational data: Euro-Atlantic jet regimes}

To represent real atmospheric data, two so-called reanalysis data sets are used. Actual observational data, whether from stations or satellite, are always unevenly distributed in time and space and therefore contain gaps. Reanalysis data fills in these gaps by blending observations with short-range weather forecasts using data assimilation methods. Two such reanalysis data sets are used here: ERA20C \cite{Poli2016}, which covers the period 1900-2010, and ERA-Interim \cite{Dee2011}, which covers the period 1979-2015. Because the period prior to 1979 suffers from a lack of satellite data, the ERA-Interim data set is generally considered more reliable. However, for the purpose of this paper, where we are looking to detect fine structure in phase space, we present results using the longer ERA20C data set only. ERA-Interim data was found to produce qualitatively similar results, and so is not shown.

The general suitability of ERA20C for regime-based studies has been commented on in previous studies \cite{Parker2019, Strommen2020}, and essentially relies on the fact that there is a long and consistent record of surface observations in the Euro-Atlantic sector, which will be our area of interest. The existence and properties of regimes in the wintertime Euro-Atlantic circulation has been extensively studied, either through the prism of pressure fields, typically geopotential height at 500hPa, or winds, in the form of zonal winds at 850hPa (hereafter ua850). Studies based on pressure data \cite{Vautard1990, Michelangeli1995, Dawson2012, Dorrington2020, Falkena2020} typically use clustering algorithms to classify distinct regimes. On the other hand, wind data is usually processed more directly in order to capture the variability of the North Atlantic eddy-driven jetstream, a relatively coherent stream of zonal winds. By measuring the location of the maximum wind-speed of the jet, one can define the latitude of the jet on any given day: the histogram of this jet-latitude index is visibly and robustly trimodal, suggesting the existence of three distinct regimes \cite{Woollings2010b}. The differences between these two perspectives, which would a priori be expected to be equivalent, can be reconciled by taking into account the added variability coming from the speed of the jet, after which both pressure and wind data suggest three very robust regimes \cite{Madonna2017, Strommen2020}. Applications to predictability have been studied in both contexts, see, e.g., \cite{Cassou2008} and \cite{Strommen2020}.

In this paper we will be focused on seeing how our framework views these three jet regimes, and so define a data set we will refer to as `JetLat'. This will be a 3-dimensional data set consisting of the daily jet latitude, and the daily values of the first and second principal components of ua850 anomalies. Data is always restricted to the North Atlantic region, defined by 15N-75N, 300E-360E, and the winter season December-January-February (DJF). The jet latitude was computed using the methodology of \cite{Parker2019}, which also includes a discussion of the jet in ERA20C.

We note that we choose  to use a data set which explicitly contains the jet latitude, already known to be multimodal,  because we wish to validate our methodology against known regime systems before applying it to less well-understood contexts. The question of locating these jet regimes using unprocessed data (i.e., data not containing any prior knowledge of the jet-latitude index) will be discussed in the conclusions, in Section \ref{sec:conclusions}. The results of applying our methodology to pressure data will be discussed in Section \ref{sec:results}.

\section{Persistent homology for dynamical systems}
\label{sec:background}

Over the last 20 years, methods from the mathematical area of topology have been increasingly used to study data analysis problems. In this section we discuss how some of these methods can be used to study  dynamical systems.

\begin{figure}[h!]
    \centering
    (a)\includegraphics[scale=0.15]{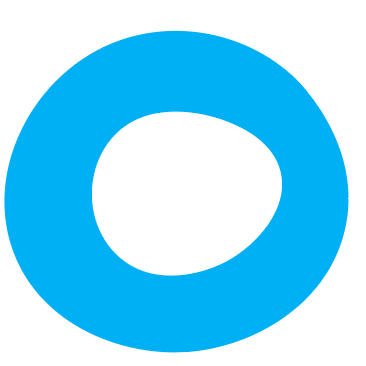} \hspace{0.5cm}  
    (b)\includegraphics[scale=0.2]{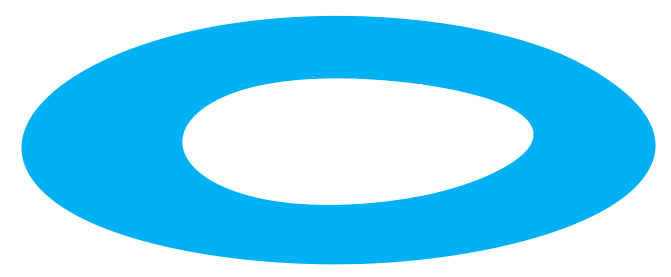}  \hspace{0.5cm} 
    (c) \includegraphics[scale=0.15]{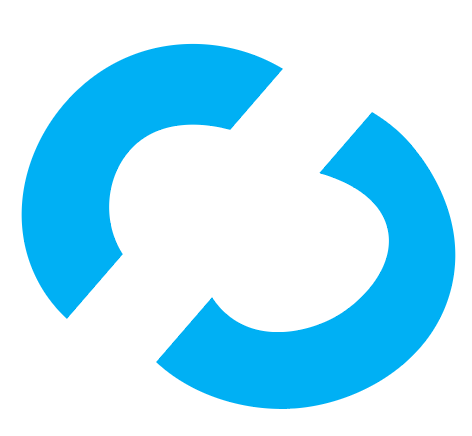}
    \caption{(a) An annulus, and (b) a shape obtained by continuously deforming the annulus, which has the same number of pieces (components) and loops as the annulus. (c) A space obtained from the annulus by a deformation that is not continuous, and thus with a different number of pieces and loops. }
    \label{F:annulus}
\end{figure}

In topology one is  interested in studying properties of shapes that do not change when one continuously deforms the shape, for instance when one squeezes or bends it.
If one considers an annulus as in Figure~\ref{F:annulus}, then no matter how the annulus is bent or stretched, it will still be composed of one piece, and have one loop. One says that the number of pieces and loops of a shape are topological invariants. On the other hand, deformations that are not allowed include cutting or gluing. If one was to cut the annulus in half, as illustrated in Figure~\ref{F:annulus}, one would break it into separate pieces with no loops. 
One can think of these invariants as giving a very coarse description of the shape of a space or of data.

In particular,  topological invariants often do not depend on the choice of of parametrisation, coordinates, or ambient dimension, and thus they are independent of many choices introduced during preprocessing steps. This is an aspect that is crucial in our work.  We caution the reader that  the  interpretation of the topological invariants,  thus what information they capture, and whether they are coarse or not, is context-specific, and depends on the specific application.

There are different ways to use topology to study data, see the survey \cite{carlsson08} for an overview.  Persistent homology, which is the method that we use in our work, is one of the standard techniques and has been very successful in many applications. In the remaining part of this section we first introduce persistent homology, and we then explain how it can be used to study the specific types of dynamical systems that we study in our work. Here we  provide an informal description, while  we provide rigorous definitions in Appendix \ref{SS:definition of barcode}.  Readers keen for even more extensive background may consult \cite{otter2017roadmap} and references therein.

\begin{figure}[h!]
\includegraphics[scale=0.35]{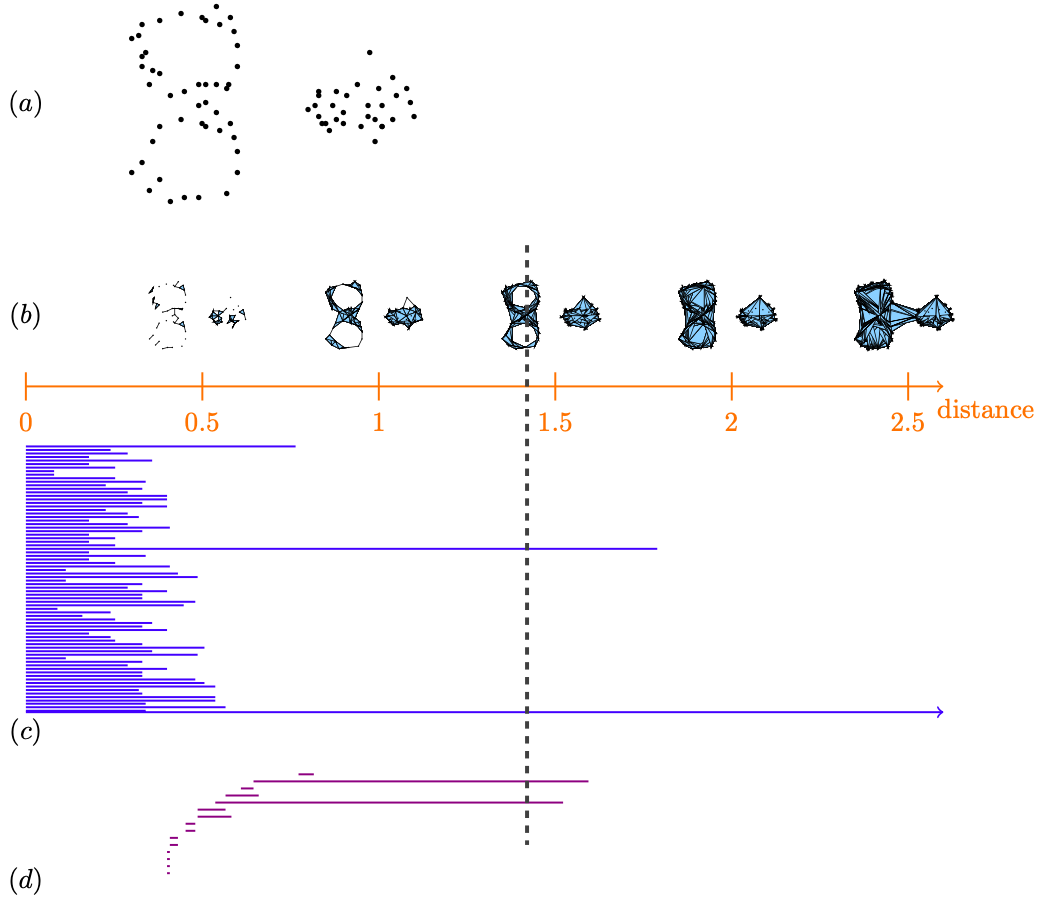}
\caption{(a) A set of points lying on a plane, with similarity given by proximity in the Euclidean distance. (b) A filtration of nested spaces, called a `Vietoris--Rips complex', obtained by connecting points within a certain distance by an edge, and filling in resulting triangles. Barcodes describing how long the (c) components and (d) holes persist in the filtration. We illustrate using a gray vertical dashed line, how we can read off information from the barcodes:  at filtration value $1.4$ there are two components, since the dashed gray line intersects two blue intervals in the barcode corresponding to the components, as well as two holes, since the gray line intersects two purple intervals in the barcode for the holes.}\label{F:example PH}
\end{figure}

\subsection{Persistent homology: informal overview}
\label{sec:pers_hom}

Given experimental data composed of points or vectors representing measurements, together with a measure of similarity (e.g., given by proximity, or correlation), in persistent homology one considers a thickening of the data set at increasing similarity scales, see Figure~\ref{F:example PH} for an example. This process yields a nested sequence of increasingly thickened spaces, which are collectively called a `filtration'. One then analyses the evolution (so-called `persistence') of the number of components, holes, voids, and higher-dimensional holes (which we call `topological features') across the filtration.

The barcode is an algebraic invariant that summarises how the topological features evolve across the filtration: the left endpoint of an interval in the barcode (the horizontal lines in Figure~\ref{F:example PH}(c) and (d)) represents the birth of a feature (the smallest distance value at which a component or hole appears in the filtration), while its right endpoint, roughly, represents the death of the same feature (the smallest distance value at which two components merge or a hole is filled in): the difference between the death and birth is referred to as the lifetime of the feature. When a feature is still  `alive' at the largest thickening scale that one considers, the lifetime
interval is by convention set as an infinite interval. For instance, we can read off from  Figure~\ref{F:example PH}(c) that there are two components that have significantly longer lifetimes than the others (corresponding to the cluster of points forming a figure-eight on the left of the figure, and the cluster of remaining points on the right), while from  Figure~\ref{F:example PH}(d) we can infer that there are two holes that live much longer than the others, which correspond to  the two holes in the figure-eight cluster. We provide a rigorous definition of holes and barcodes in Appendix \ref{SS:definition of barcode}.

The interpretation of the intervals in the barcode that we have given here is only one of the possible applications of persistent homology to the study of data. In other types of applications, it might be the intervals of a certain length, and not necessarily the longest ones,  that encode significant information, see for instance, \cite{Bendich2016,bubenik2020persistent}. In particular, the interpretation of the barcode is application specific. We discuss how we interpret the barcode in our work in more detail in Section \ref{sec:significance_test}.

\subsection{Computational complexity of PH}
\label{sec:complexity}
The theory behind (one-parameter) persistent homology is well-understood, and amounts to standard linear algebra. Conversely, the computation of the barcode is  expensive, since the computational complexity can grow exponentially in the size of the input data, in the worst case. To sidestep such difficulties, in this work we use optimised algorithms, and sparsification techniques, see also Section \ref{sec:parameters}. We refer the reader to the survey \cite{otter2017roadmap} for a  detailed discussion of the computational complexity of the main persistent homology algorithms.

The types of filtered spaces that we consider here rely on the computation of  distances between points, and such distances have computational cost that is, in the worst case, linear in the dimension of the ambient space. Once the distances are computed, the computational complexity of persistent homology of these filtered spaces depends on the size of the input data, and, thus, on the number of measurements, but not on the dimension of the ambient space. Thus, a consequence that this has for our work is that adding  variables to our models increases the computational cost only  by a linear function of the dimension of the ambient space. This is one of the reasons that persistent homology is so effective for the study of dynamical systems, especially in atmospheric science where spaces are heavily undersampled and the dimensionality of the phase space is often orders of magnitudes larger than the number of measurements.

\subsection{Optimal representatives of cycles}\label{SS:repr cycles}

\begin{figure}[ht]
    \centering
    \includegraphics[scale=0.25]{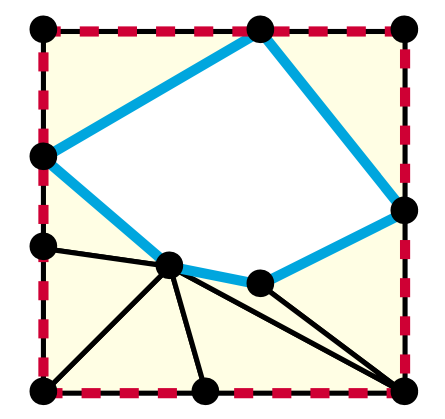}
    \caption{Two $1$-cycles that are representatives for the same hole: (cyan, solid) an optimal (i.e., minimal)  representative, and (red, dashed) a not optimal one.
    }\label{F: cycles}
    \label{fig:my_label}
\end{figure}

Given an interval in the barcode describing the lifetime of a component or hole, we are interested in studying the points in the data that correspond to such a component or hole. Such points are called `representatives' for holes in  dimension $0$ (i.e., components) and in dimension $1$ (i.e., holes). We refer the reader to Appendix \ref{SS:definition of barcode}, and in particular Definition \ref{D:simplicial homology} and the preceding paragraph for a definition of $p$-dimensional holes and $p$-cycles. 

Ideally, we want to be able to choose representatives that are easily interpretable from a geometric point of view. For instance, we might want representatives for $1$-dimensional holes to have minimal length in a suitable sense, see the illustration  in Figure~\ref{F: cycles}. 

Thus, we are interested in representatives that satisfy some minimality condition:  for holes we 
 compute optimal representatives \cite{DHM19} using the software Persloop \cite{persloop}, while for components, we use representatives to find all the points in a component.
 We note that finding optimal representatives for holes is a challenging problem; the software Persloop implements an algorithm that gives a heuristic approximation for $1$-cycles in $3$D, but which might fail to give meaningful $1$-cycles on higher dimensional data sets.

\subsection{Multiparameter persistent homology}

\begin{figure}[ht]
    \centering
    \includegraphics[scale=1]{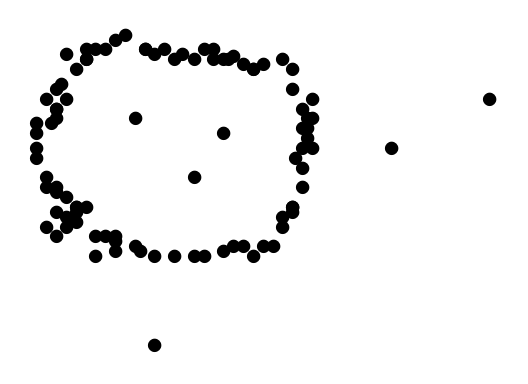}
    \caption{A finite set of points in $\mathbb{R}^2$ for which a distance-based filtrations might fail to capture interesting topological information.}
    \label{fig:bifiltration pointcloud}
\end{figure}

\begin{figure}[ht]
    \centering
    \includegraphics[scale=1]{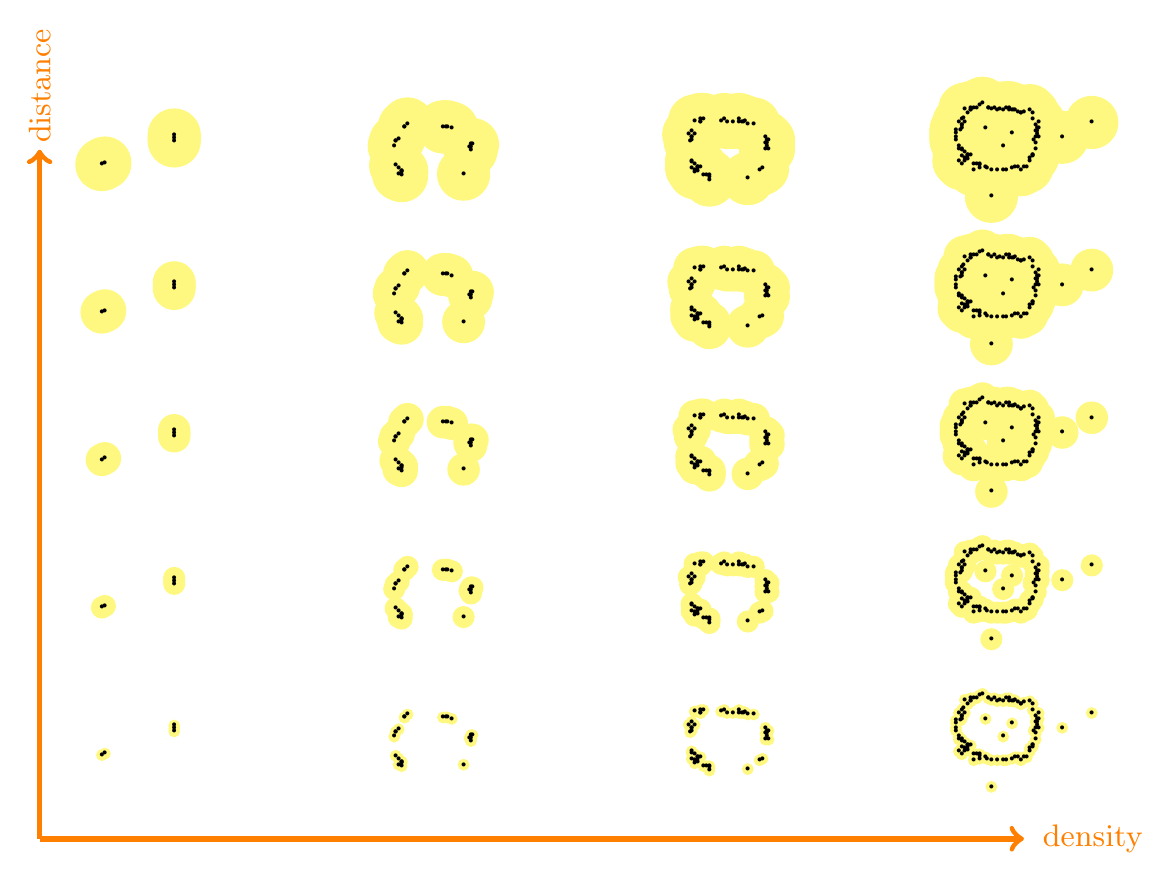}
    \caption{A bifiltration obtained by decreasing density and increasing distance:  given the finite set $X$ of points in $\mathbb{R}^2$ illustrated in Figure \ref{fig:bifiltration pointcloud}, and a density estimation,  we consider subsets $X'\subset X$ of points having density above a certain threshold. For each subset $X'$ of points we then construct a distance-based filtration by taking balls with increasing radii centered at the points.}
    \label{fig:bifiltration}
\end{figure}

In many application problems, one might wish to  study filtrations that depend on more than one parameter. For instance, consider the point cloud in $\mathbb{R}^2$ in Figure~\ref{fig:bifiltration pointcloud}. If one were to consider only the points belonging to higher-density regions, one could associate to these points a distance-based filtration, as illustrated in Figure \ref{F:example PH} and discussed in Section \ref{sec:pers_hom}. Then, by computing the persistent homology  of such a  filtration, one could read off from the barcodes  that the point cloud has a long-lived component, and a long-lived hole. For such a data set, it might be difficult in practice to choose the right density value, and therefore  one would ideally wish to consider point clouds thresholded at all possible density values, thus obtaining a bifiltration, as illustrated in Figure \ref{fig:bifiltration}.

The theory of persistent homology does not generalise to filtrations that depend on more than one parameter. In particular, there is no generalisation of the barcode, as described in Section \ref{sec:pers_hom} and illustrated in Figure \ref{F:example PH}, for multifiltrations. Finding appropriate ways to quantify the `persistence' of topological invariants, such as the number of components or holes, is currently one of the most active areas of research in TDA, and several researchers have proposed invariants that are computable, and capture in an appropriate sense what it means for topological features to be `persistent', see, for instance, \cite{HOST, LW15,V20}. In Section \ref{sec:rivet} we discuss one such approach.

\subsubsection{Barcodes along one-dimensional subspaces}
\label{sec:rivet}
In one approach to defining invariants for multiparameter persistence that are suitable for applications, researchers study ways to restrict a bifiltration, such as the one in Figure \ref{fig:bifiltration}, to one-dimensional subspaces, and  then study  barcodes along such restrictions \cite{LW15,BCF+08}.

\begin{figure}[ht]
    \centering
    \includegraphics[scale=1]{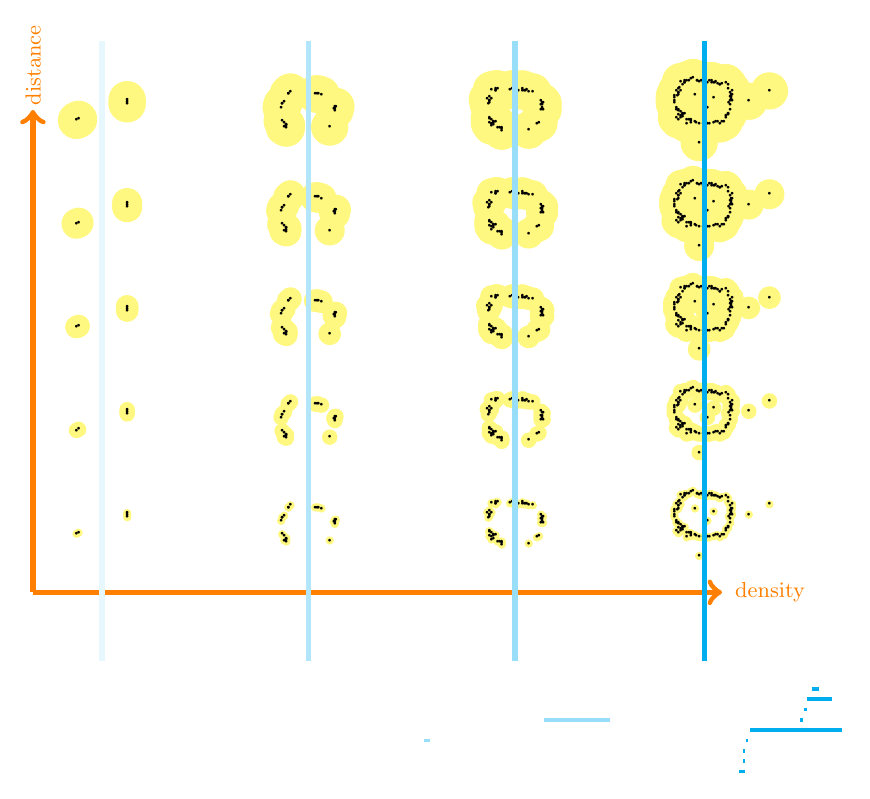}
    \caption{Barcodes (the collection of blue intervals at the bottom of the figure) for holes along restrictions to vertical lines of the bifiltration from Figure \ref{fig:bifiltration}. The barcodes, for the holes, for the first two lines are empty due to the lack of any holes. We note that the barcodes for the components, which we are not depicting here, are not empty.}
    \label{fig:bifiltration slice}
\end{figure}

As illustrated in Figure \ref{fig:bifiltration slice}, restricting oneself to points up to  a specific density threshold amounts to considering a filtration of spaces along a vertical line in the bifiltration. By studying persistent homology of this filtration, we are thus computing the barcode of the restriction of the bifiltration along this line. More generally, one could consider lines with any slope in the $2$-parameter space, and then compute the barcode of the restriction of the bifiltration along this line. It is known that this process is robust in an appropriate sense only for lines having positive slope, see the discussion in  \cite[Section 1.5]{LW15}. In particular, here, if we consider filtrations for different density threshold levels, we might observe intervals suddenly appearing or disappearing in the corresponding barcodes. 

Lesnick and Wright implemented their methods \cite{LW15} in the software package RIVET \cite{rivet}, which is currently the only existing software package for the computation of multiparameter persistent homology.  
Unfortunately, the current implementation in RIVET is not memory-efficient enough for the types of data sets that we study in our work, since if one is interested in computing barcodes to study the lifetime of  loops, the software can only handle data sets of a few hundred points. Thus, one main direction that we plan to pursue in future work, is to work on optimisations of the computations implemented in RIVET. In particular, in future work we plan to compute barcodes along restriction to lines with positive slopes, to obtain a method that is robust.


\subsection{Bifiltrations for dynamical systems}
\label{sec:bifiltrations}

\begin{figure}[ht]
    \centering
    \includegraphics[scale=0.8]{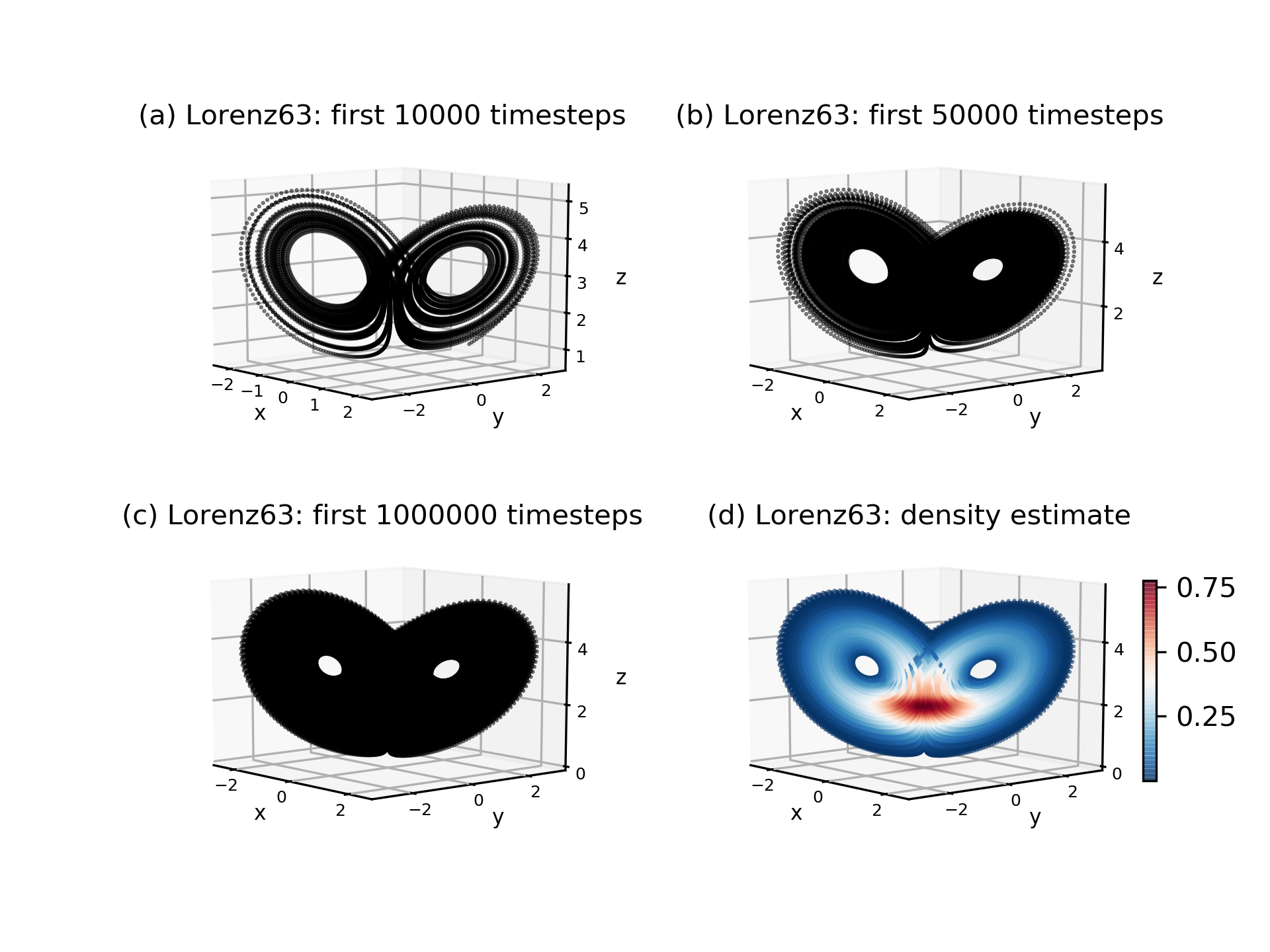}
    \caption{In (a)-(c): the Lorenz `63 system visualised using 10000, 50000 and 1000000 timesteps respectively. In (d), the Lorenz `63 system using 1000000 timesteps, with colours representing the density, as measured with the kernel density estimator.}
    \label{fig:lorenz_holes_evolution}
\end{figure}

The need to consider not just a filtration of distances, as in the standard method of one-parameter persistent homology, but a bifiltration of distance and density, can be motivated here in two ways. Firstly, and most fundamentally, the dynamical systems we are interested in are always \emph{continuous}, and so no two regions on the attractor can be fully disconnected from each other. In fact, the connectedness of the attractor of a continuous dynamical system can be proved mathematically, given a suitable definition of `attractor' \cite{Gobbino1997}, implying that persistent homology will never detect more than one long-lived connected component from a generic sample of the system. The second reason can be understood by considering the Lorenz `63 system. In Figure~\ref{fig:lorenz_holes_evolution} we demonstrate a particular feature of the system, namely that the size of the two iconic holes become smaller as one increases the sample size. This implies, somewhat paradoxically, that topological features may become harder to detect the more points you have. If the size of the features becomes comparable to the distance between consecutive points, then these features may, practically speaking, become impossible to detect computationally. These two observations suggest that a naive application of persistent homology to a continuous dynamical system may easily fail to detect both long-lived connected components and long-lived holes.

The basic underlying problem is that in one-parameter persistent homology one computes a filtration by increasing Euclidean distances between the points, ignoring any variations in density. However, the regimes classically identified with clustering methods typically correspond to regions of above-average density, suggesting that the connected components we are interested in should be relative to density. Furthermore, in the Lorenz `63 system, the reason any generic sample of the attractor yields visually clear holes is the fact that the regions of phase space close to the centre of the holes, i.e., near the fixed points, are very low density regions. Therefore, the holes in the system are only identified in data with respect to some chosen density threshold.

Persistent homology provides a solution to the problem of choosing a density threshold at which to study points: instead of trying to estimate the best value for the density parameter, we consider a {\em bifiltration} of distance and density. An example of such a bifiltration was given in Figure \ref{fig:bifiltration}. We note that the bifiltration, and hence the corresponding topological properties that one observes, depend on a choice of density estimation function.

As a final remark, it may be possible to achieve good results by extending the filtration to other measures besides density. In particular, our tests suggest that using phase space velocities can, in some situations, be equally useful: prior knowledge of the system of interest might inform more particular choices. We note that the use of a computationally costly measure of density may, to an extent, offset some of the computational gains of persistent homology discussed in Section \ref{sec:complexity}. The use of computationally cheaper measures, such as phase space velocities, may therefore be preferable for larger applications. For the context of this paper, however, we will only consider density.

\section{Computational methodology}
\label{sec:methodology}

We now describe the full algorithm that we perform to analyse a given data-set sample. The basic method is the following:

\begin{itemize}
    \item[(1)] Normalise each dimension in the data set to have unit variance.
    \item[(2)] Estimate the density at every point in the normalised data-set sample $D$.
    \item[(3)] Pick a percentage threshold $P\%$. Select the sub-sample $D_P$ defined by the upper $P$th density percentile of $D$, i.e. the $P\%$ densest points of $D$.
    \item[(4)] Compute persistent homology for $D_P$ and extract the topological features of interest: birth/death times for each cycle detected; the points belonging to each of the five longest-lived connected components; a topological representative of each of the five longest lived loops.
    \item[(5)] Repeat for values of $P$ ranging from $10\%, 20\%, \ldots, 100\%$ and examine the features that appear in the resulting bifiltration.
\end{itemize}

The essentially arbitrary choice to only show the 5 longest-lived cycles was made to ensure visual tidiness in plots. For the data sets considered here, no important information is lost by this restriction, though this will of course not be true in general. Note that depending on the choice of parameters (see Section \ref{sec:parameters}), less than 5 cycles may be found.  We also note that the normalisation in step 1 is important to ensure that interesting structure is not missed purely by virtue of existing along a direction in phase space with smaller magnitudes, such as loops that appear as `squashed' ellipses in the raw data. Finally, in all our examples we use the percentage thresholds $10\%, 20\%, \ldots, 100\%$. This choice suffices to recover the main features of the systems we consider: applications to unknown dynamics may require finer thresholds. The methodology presented here is therefore best understood as a technique that can be flexibly adapted to new situations, much like classical one-parameter persistent homology. Further details on the other steps now follow.

\subsection{Density estimation}
\label{sec:density_estimation}

The primary method used was a kernel density estimator (KDE) with a Gaussian kernel \cite{Marron2007}, computed using inbuilt functions of the scipy python package \cite{2020SciPy-NMeth}, where we used the default option of Scott's Rule to determine the bandwidth. The bandwidth determines the minimal spatial scale of the features we compute. We are primarily interested in ignoring features occurring at scales comparable to the average distance between points at consecutive timesteps, which we will informally refer to as the `grid-scale' of the data set. Using a KDE has two clear advantages for topological applications. Firstly, it produces smooth estimates, which avoids potential issues whereby outlier points remain even after a severe density threshold has been applied. Such outliers will often appear as spurious long-lived connected components, effectively just adding noise to the analysis. Secondly, KDE's are well-suited to represent multimodality, a key feature we want to capture.

A second, cruder method was also tested, which involved directly binning the space and counting the datapoints in each bin. To facilitate the computations, this density estimate was carried out in the space spanned by the first three EOFs, under the assumption that the resulting estimate would be accurate for the scales we were interested in studying. A fixed number of $160^d$ bins were used in each case, where $d$ is the dimension of the data set.

In our results, the KDE produced good results in all cases except for the CdV system. As will be shown, the CdV system exhibits some very fine-scale structure in the form of `thin', low-density loops that emerge within a larger, more chaotically-inhabited, low-density region. The Gaussian KDE we used was found to smear away a lot of this structure, while the direct binning method picked out these features easily. In the other data sets, both the KDE and direct binning methods produce qualitatively similar results, but the KDE exhibits a notably smoother estimate, as expected. For this reason, results obtained using the KDE are shown for all data sets except CdV, where the results obtained with direct binning are shown instead. It would clearly be of interest to address the question of whether a more appropriate density estimation method (e.g. choice of bandwidth) might yield good results in all cases, but this is left for future work.

\subsection{Computation of persistent homology and representative cycles}
\label{sec:parameters}

In the present work, we compute persistent homology using the Vietoris--Rips complex (see Figure \ref{F:example PH}) and the python package Gudhi \cite{gudhi}. Gudhi takes as input both the data set and several user-specified input parameters, the choice of which we now outline. 

\begin{itemize}
    \item \verb|max_edge|: This parameter  determines the maximal distance threshold to consider in the filtration. Setting this as the maximal distance between any two points in the data set guarantees that the filtration terminates (i.e., ends with a single connected component at the end), so this parameter can always be chosen in a principled manner. Because all our data sets were normalised, we were able to set \verb|max_edge| $=5.0$ for all data sets.
    \item \verb|min_pers|: This parameter determines the minimal lifespan that a computed homological cycle needs to attain in order to be included in the final output from Gudhi. The choice of this parameter therefore determines the scales of the topological features one wants to consider, similar to the choice of bandwidth in the density filtration (cf. Section \ref{sec:density_estimation}). While prior knowledge of the spatial scales of the system can be used to inform the choice of this parameter, the only downside in setting this parameter as very small is an associated increase in computational cost. Because all our data sets are normalised, we found that a parameter choice between $0.15$ and $0.50$ gave good answers at low cost for all data sets. The higher value was used for CdV, as the main features there exist at higher scales, while for systems like JetLat, with subtler behaviour, the smaller value was used.
    \item \verb|sparse|: This parameter is internal to Gudhi's algorithm, and determines the extent to which the computed Rips complex is sparsened before computing persistent homology. This was set to $0.7$ for all data sets.
    \item \verb|pre_sparse|: This parameter is fed in to the Gudhi \verb|sparsify_point_set| routine, which is used to perform a preliminary sparsification of the data set prior to carrying out computations. The routine is built to sparsify data sets in a way which does not change the topology, e.g., by replacing densely connected regions with a sparser set of points covering the same region. Because our data sets are always filtered by density prior to computation, such a sparsification has no impact on our results, but allows the computations to be sped up significantly. In fact, for large data sets with a time dimension exceeding 30000 timesteps, computations would typically run out of memory and crash. Setting an appropriate value of \verb|pre_sparse|, which greatly reduces the number of points, was therefore crucial. In practice, we set this value as the smallest positive number which would allow the computations to finish at a reasonable rate. A value of $0.05$ was found to be suitable for Lorenz `63, Lorenz `96, while $0.005$ worked best for CdV. For the JetLat data set, where the total number of time-steps available is only around 10000, this sparsification step was not necessary and hence not carried out.
\end{itemize}

After computing the filtration and homology at a given density threshold, the five longest-lived components and loops were identified. Explicitly determining the points belonging to each connected component can be done easily using output from Gudhi, which gives the full filtration. By keeping track of which points are linked up as the filtration radius grows, basic python code suffices to determine all the components; the code used is freely available online (see the Data Availability Statement).

We note that obtaining a representative cycle of loops is significantly harder, as discussed in Section \ref{SS:repr cycles}.

\subsection{Sensitivity to parameter choices}

Several tests were carried out to determine the sensitivity of our results to the parameter choices described in the previous section. A selection of density thresholds for the different data sets were chosen at random, and standard birth/death plots produced using Gudhi for the resulting filtered data sets. It was found that the qualitative features of these birth/death plots did not appreciably change in response to mild perturbations of the parameters, implying that the basic topological features, as summarized in our bifiltration plots, do not depend sensitively on our choices. The size and location of connected components was also found to be largely insensitive to such parameter changes.

On the other hand, the representatives of loops, as computed with PersLoop, were found to exhibit sensitive dependence, in particular on the \verb|pre_sparse| parameter. A small perturbation of this parameter would often lead to the software not terminating properly, or producing a very different representative loop. A similar phenomenon was observed when keeping parameters fixed, but changing other aspects of pre-processing, such as the choice of density filtering or the use of EOF data versus raw data. The reader should therefore be cautioned that the representative cycles we show in our plots are not to be viewed as reliable output from a stable algorithm. Rather, they are included to demonstrate that the topological features seen in our bifiltration plots can, in principle, be visualised in the data itself, and really do correspond to the features one expects.

\subsection{Significance testing and topological non-triviality}
\label{sec:significance_test}

In this paper, we take the stance that Gaussian distributions should be viewed as  having no interesting topological structure, in the sense of persistent homology, and no meaningful regimes. Therefore, in order to assess whether features identified in our bifiltration methodology are more than just sampling noise, we implemented the following procedure. First we draw 10000 random samples from a three-dimensional Gaussian distribution with unit variance. Secondly, we run this through our methodology described at the beginning of Section \ref{sec:methodology}. The maximal lifespans of both the connected components and loops obtained at any of the density thresholds were kept: the whole procedure is then repeated ten times and the maximal lifespan obtained across all random draws is used as a measure of noise. Specifically, features with a lifespan close to this value, of around $0.4$, are likely to be noise coming from grid-scale sampling variability, while features with a lifespan greatly exceeding this are likely to be indicative of significant non-trivial homology. In the context of this paper, we therefore define a dynamical system to have non-trivial topological structure if and only if its distance-density bifiltration produces cycles with lifespans exceeding that expected from Gaussian noise (i.e., lifespans exceeding 0.4, in the case where the dimensions of the systems have been normalised).

When carrying out this procedure, the connected components in Gaussian samples containing 3 or less points were not included, because one or two big outlier points can easily produce very long-lived `components'. For consistency, components with 3 or less points that are detected in any data set are always clearly marked in plots. Note that such outliers can also occur for filtered data unless \verb|min_pers| is large, since there may randomly be some points fractionally closer to each other than any other points.

Finally, we note that because all our data sets are normalised prior to computing homology, the unit variance Gaussian offers an appropriate comparison for all the data sets that we consider.

\section{Results}
\label{sec:results}

For each data set, we now produce a standard bifiltration plot summarising the lifespans of persistent cycles across a range of density thresholds. In addition, to visualise these topological features, particular density thresholds are hand-picked for each data set and plotted, together with a visualisation of either the connected components or the representatives of loops present at that threshold.

\subsection{The Gaussian}

\begin{figure}[ht]
    \centering
    \includegraphics[scale=0.7]{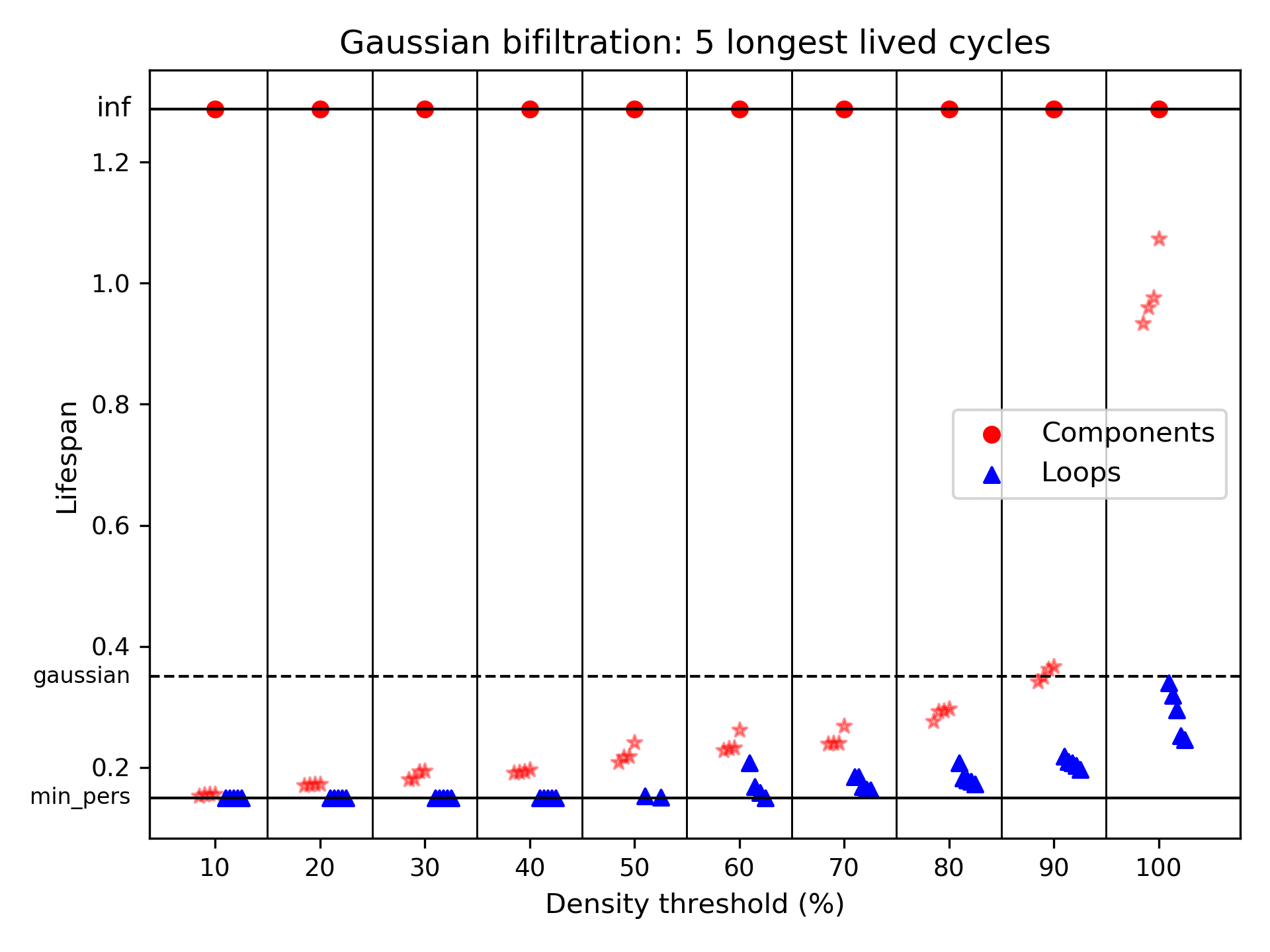}
    \cprotect\caption{A distance-density bifiltration of a unit variance Gaussian distribution. For each density threshold on the $x$-axis, the lifespan of the 5 longest-lived connected components (red dots if the component contains more than 3 points: red stars otherwise) and 5 longest-lived loops (blue triangles) are plotted. The stippled line shows the largest lifespan obtained across multiple Gaussian samples. The meaning of the \verb|min_pers| parameter is explained in Section \ref{sec:parameters}.}
    \label{fig:gaussian_bifiltration}
\end{figure}

\begin{figure}[ht]
    \centering
    \includegraphics[scale=0.8]{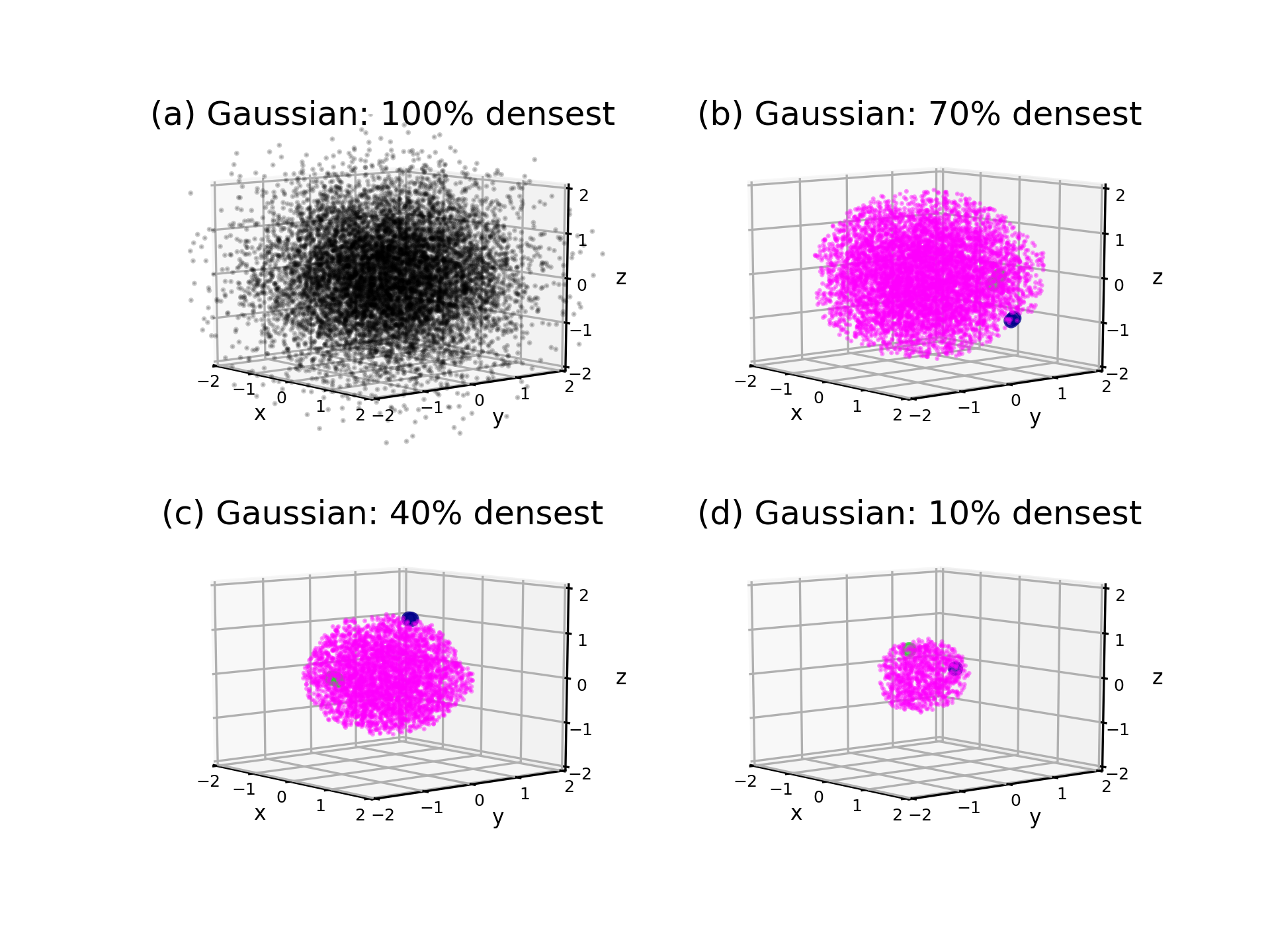}
    \caption{In (a): a random sample of a unit variance Gaussian distribution. In (b): the 70\% densest points of the sample; (c): the 40\% densest points; (d): the 10\% densest points. In (b)-(d), points are coloured according to the connected component they live in: longest-lived (red colour), 2nd longest lived (blue colour) and 3rd longest lived (green colour). Points belonging to components with 3 or less points have been made larger to aid visualisation.}
    \label{fig:gaussian_thresholds}
\end{figure}

As explained in Section \ref{sec:significance_test}, results from the unit variance Gaussian distribution are used to estimate the significance of features obtained for all other data sets, since any definition of regimes should exclude the Gaussian from having any. We therefore first present results for a randomly drawn sample of 10000 points from such a distribution. These are shown in Figure \ref{fig:gaussian_bifiltration}. As expected, no non-trivial topological features are detected in this data set, with each density threshold exhibiting only a single connected component (the red dot at infinity) and some spurious outliers (the red stars) at the `grid-scale'. The loops found (blue triangles) are all extremely close to the minimum persistence choice, implying that these were only barely registered by the algorithm and do not persist for notably longer than isolated outlier components.

The seeming change in behaviour at the 100\% threshold, where no density filtering has been applied, is due to the existence of big outliers in the raw sample. This is clearly seen in Figure \ref{fig:gaussian_thresholds}, showing the Gaussian sample at various thresholds. Because even an extremely mild density threshold immediately removes the big outliers seen in Figure \ref{fig:gaussian_thresholds}(a), the possible lifespan of small components with 3 or less points drops dramatically from the 100\% to 90\% threshold. This is also why the longest-lived loops are found at the 100\% threshold. As is clear from Figure \ref{fig:gaussian_thresholds}, these loops are just noise, and indeed any representatives of these produced by PersLoop (not shown) are visually confirmed as such. Figures \ref{fig:gaussian_thresholds}(b)-(d) also highlight the 3 longest-lived connected components at each threshold. It can be seen that this yields one component containing almost all points, and two components consisting of one or two points that simply happen to be fractionally further removed from the rest of the point mass.

These observations already confirm that our methodology correctly identifies the Gaussian as having no non-trivial topology at any density threshold. The comparison of Figure \ref{fig:gaussian_bifiltration} with the equivalent plots for other data sets, to which we now turn, will make this even clearer.

\subsection{Lorenz `63}

\begin{figure}[ht]
    \centering
    \includegraphics[scale=0.7]{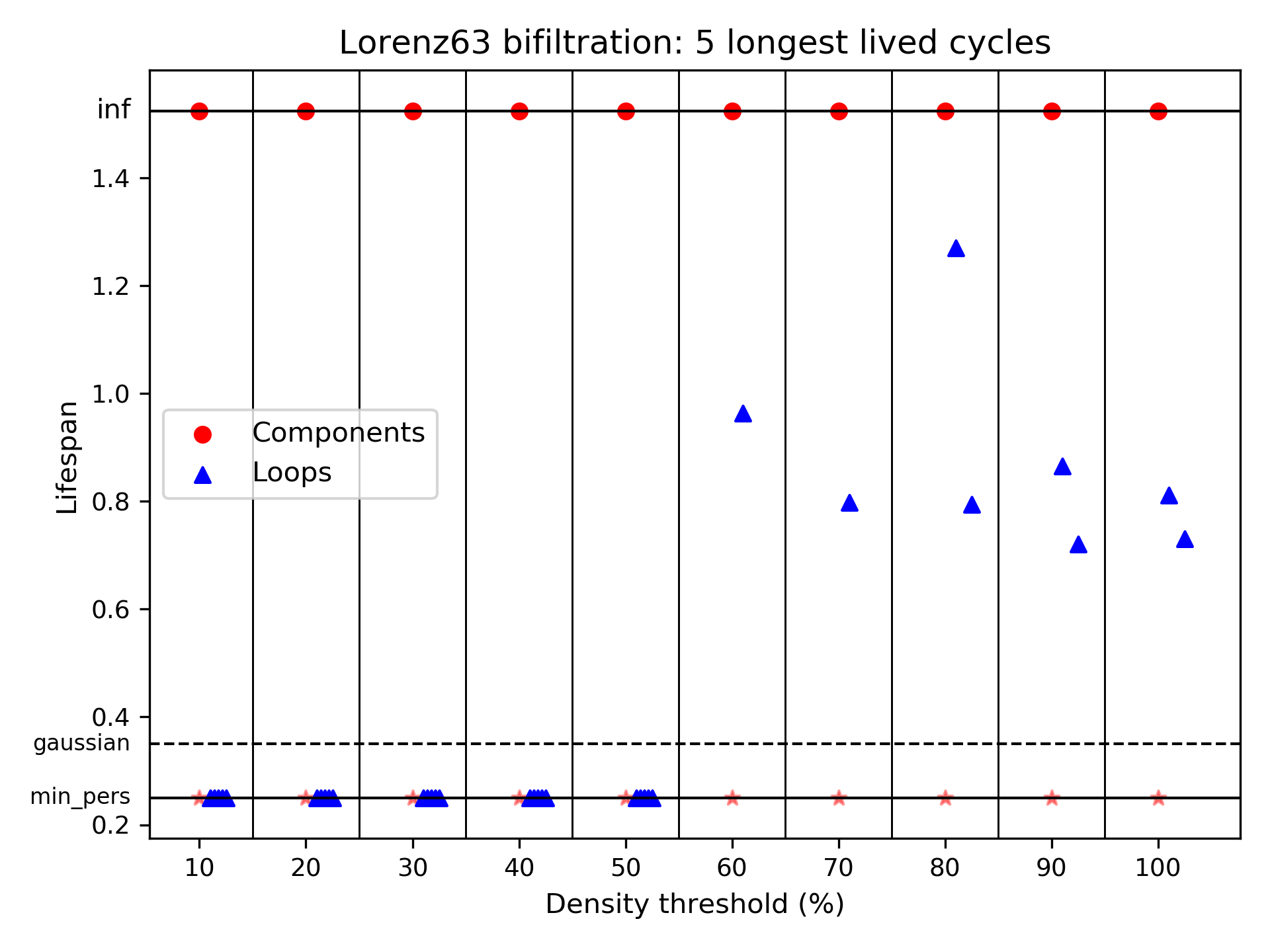}
    \cprotect\caption{A distance-density bifiltration of the Lorenz `63 system. For each density threshold on the $x$-axis, the lifespan of the 5 longest-lived connected components (red dots if the component contains more than 3 points: red stars otherwise) and 5 longest-lived loops (blue triangles) are plotted. The stippled line shows the largest lifespan expected from Gaussian noise. The meaning of the \verb|min_pers| parameter is explained in Section \ref{sec:parameters}.}
    \label{fig:lorenz_bifiltration}
\end{figure}

\begin{figure}[ht]
    \centering
    \includegraphics[scale=0.8]{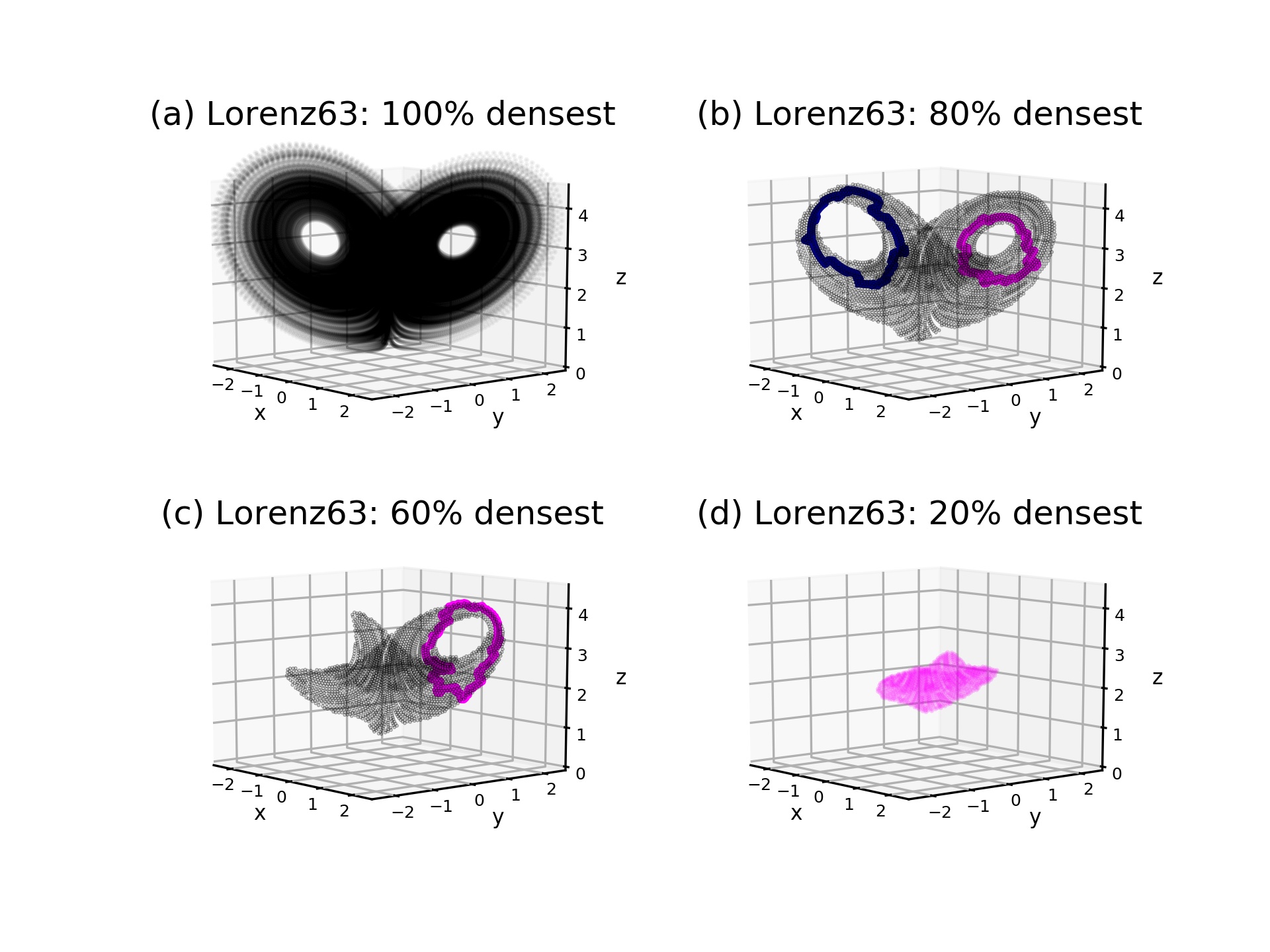}
    \caption{In (a): a long integration of the Lorenz `63 system. In (b): the 80\% densest points; (c): the 60\% densest points; (d): the 20\% densest points. In (b) and (c), representatives of the 2 (respectively 1) longest-lived loops are overlain. The longest-lived loop is always in red, the 2nd longest in blue. In (d), the longest-lived connected component is marked in red.}
    \label{fig:lorenz_thresholds}
\end{figure}

Figure \ref{fig:lorenz_bifiltration} shows the bifiltration plot of the Lorenz `63 system. This plot can be understood by reference to Figure \ref{fig:lorenz_thresholds}, which visualises the system, and the longest-lived components/loops, at different thresholds. At low density thresholds, as shown in Figure \ref{fig:lorenz_thresholds}(d), there is just one connected component, corresponding to the dense central region between the two wings. Because the density is concentrated in this area, as seen in Figure \ref{fig:lorenz_holes_evolution}(d), there is no trace of the two wings until you move to higher thresholds. At the 60\% threshold, Figure \ref{fig:lorenz_thresholds}(c), enough points are included for one of the wings to emerge, at which point an extremely long-lived hole appears in the bifiltration plot: the representative produced by PersLoop confirms that this corresponds to the right wing. At the 70\% threshold, the second wing also emerges, after which one retains two long-lived holes for all further density thresholds. Figure \ref{fig:lorenz_thresholds}(b) confirms that the two holes found by Gudhi at this point correspond to the two holes in the wings. Note that the apparent asymmetry between the two loops is due to sampling variability.

Two other points are worth observing in Figure \ref{fig:lorenz_bifiltration}. Firstly, besides the key topological features coming from the wings, all other features have lifespans at the \verb|min_pers| threshold, implying that these features exist only at or below the grid-scale of Lorenz `63. Secondly, these grid-scale features have lifespans below what is expected from a Gaussian bifiltration, demonstrating that our significance test has correctly classified these as noise. Furthermore, the lifespans of the two loops, and one connected component, greatly exceed Gaussian noise. The conclusion from our methodology is therefore that the Lorenz `63 system has two significant holes, corresponding precisely to the two classical regimes defined by the wings \cite{Palmer1994}, and is otherwise fully connected.

\subsection{Lorenz `96}

\begin{figure}[ht]
    \includegraphics[scale=0.7]{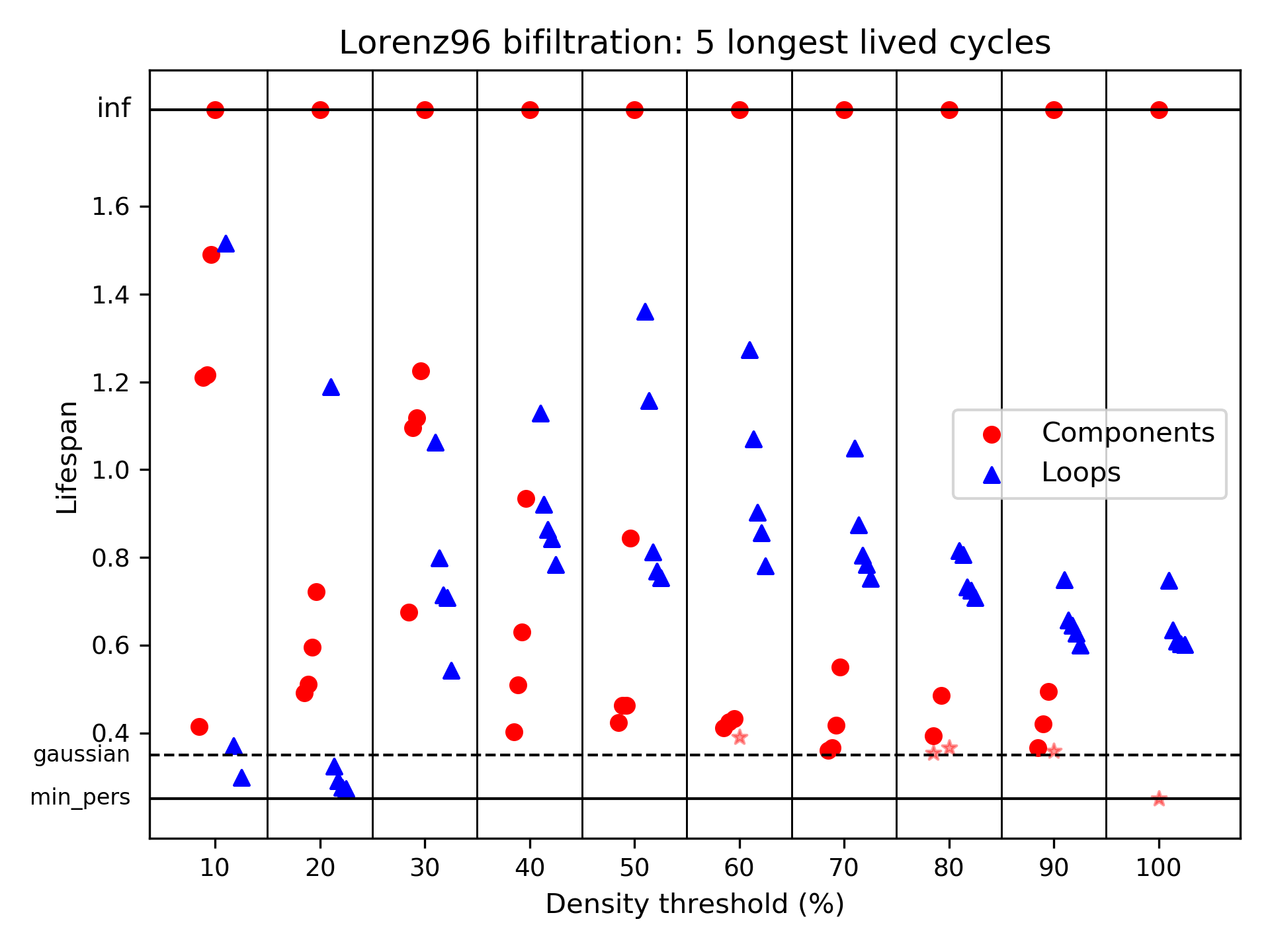}
    \cprotect\caption{A distance-density bifiltration of the Lorenz `96 system. For each density threshold on the $x$-axis, the lifespan of the 5 longest-lived connected components (red dots if the component contains more than 3 points: red stars otherwise) and 5 longest-lived loops (blue triangles) are plotted. The stippled line shows the largest lifespan expected from Gaussian noise. The meaning of the \verb|min_pers| parameter is explained in Section \ref{sec:parameters}.}
    \label{fig:lorenz96_bifiltration}
\end{figure}

\begin{figure}[ht]
    \centering
    \includegraphics[scale=0.8]{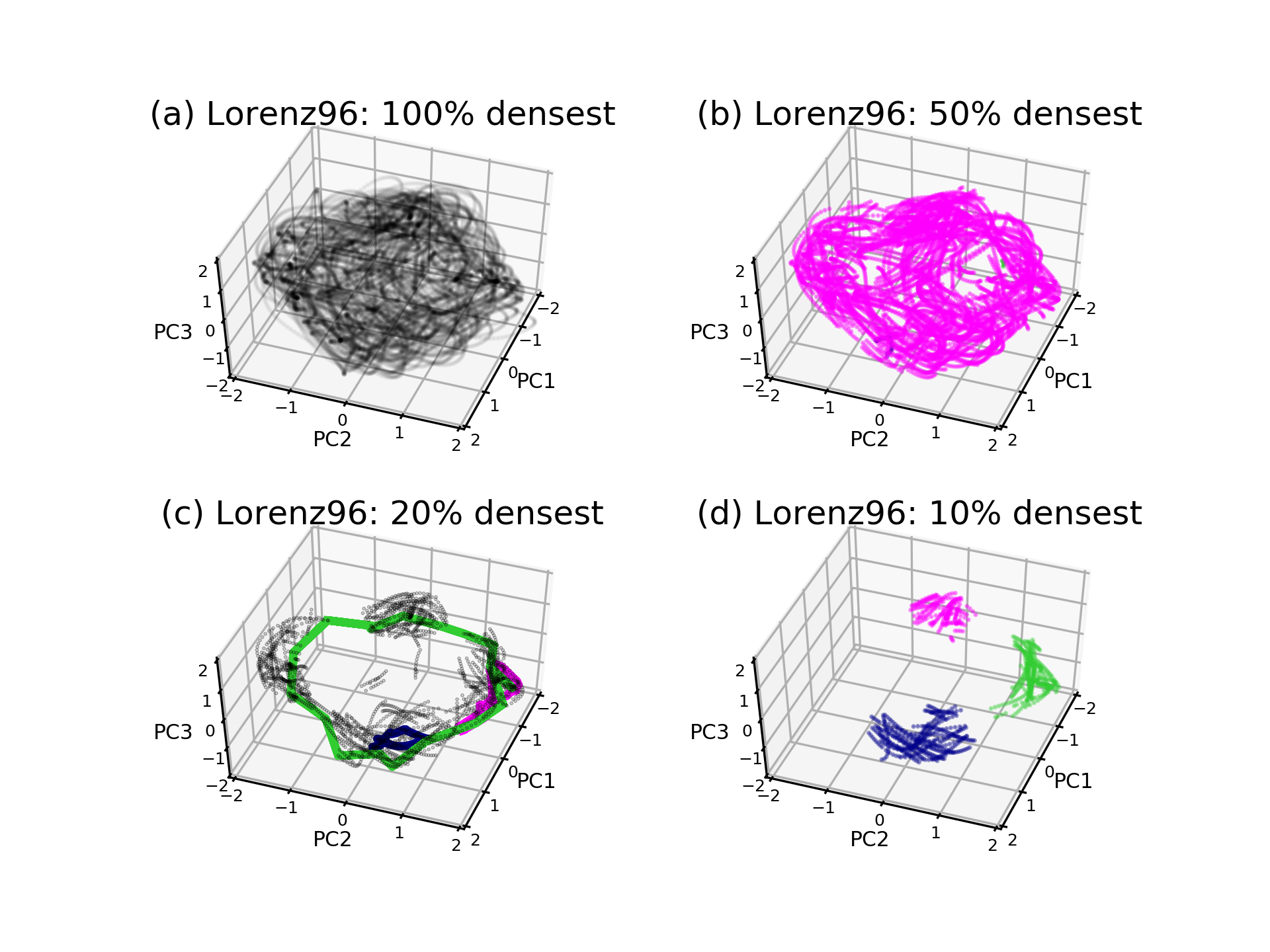}
    \caption{In (a): a long integration of the Lorenz `96 system. In (b): the 50\% densest points; (c): the 20\% densest points; (d): the 10\% densest points. In (b) and (d), representatives of the 3 longest-lived connected components are marked with colours, while in (c), the 3 longest-lived loops are overlain in colour. In all cases, the longest-lived feature is in red, the 2nd longest-lived in blue and the 3rd longest-lived in green.}
    \label{fig:lorenz96_thresholds}
\end{figure}

Figure \ref{fig:lorenz96_bifiltration} shows the bifiltration plot for the Lorenz `96 system, which suggests the existence of a considerable amount of significant topological structure. Figure \ref{fig:lorenz96_thresholds} shows some of this structure at different thresholds, though we remind the reader that because the homological computations in this case were done using a 4-dimensional EOF truncation, our 3-dimensional projections necessarily obscure some of the features. We also note that, due to the limitations of PersLoop, optimal loops were computed using the space spanned by the first three EOFs only, which also leads to some minor distortions.

The characteristic looping behaviour of the system is already visible in the unfiltered data set, Figure \ref{fig:lorenz96_thresholds}(a), reflecting the rotational symmetry in the defining equations. The looping trajectories result in regions which, after an appropriate density threshold is imposed, appear as holes in an otherwise connected space, as in Figure \ref{fig:lorenz96_thresholds}(b). The most prominent loop appearing in this manner is the one circling the full perimeter of the space, as seen in Figure \ref{fig:lorenz96_thresholds}(c). Note that PersLoop identifies this loop as the 3rd longest-lived at the 20\% threshold. The representatives found for the longest and 2nd longest-lived loops are made to look particularly spurious due to the flattening of the 4th dimension, but are either way examples of the way in which PersLoop sometimes produces representatives that are far from optimal. For very severe density thresholds, such as the 10\% threshold shown in Figure \ref{fig:lorenz96_thresholds}(d), the data set splits up into distinct components, implying significant local variations in density across the attractor.

\begin{figure}[ht]
    \centering
    \includegraphics[scale=0.7]{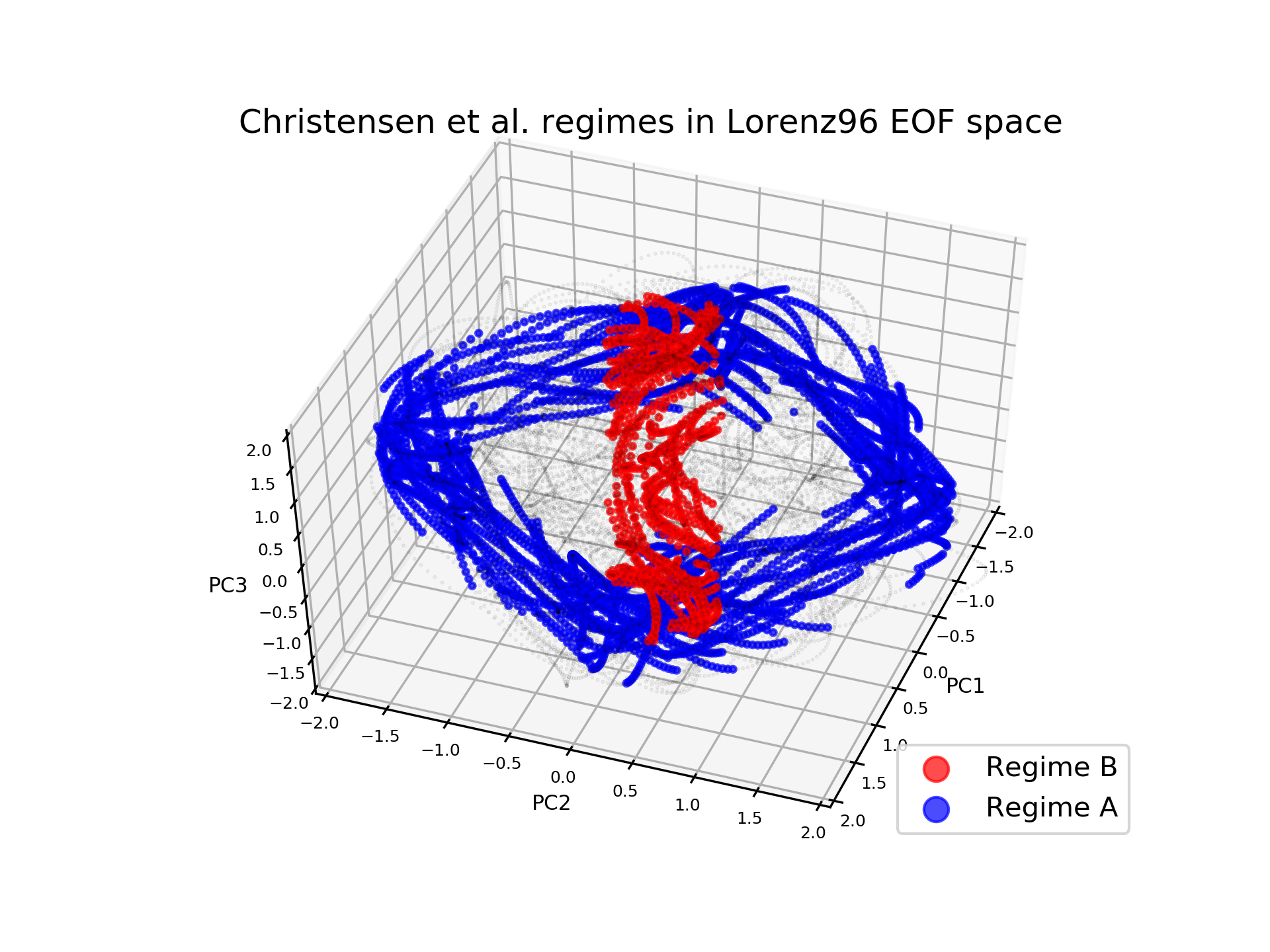}
    \caption{The two regimes $A$  (red) and $B$ (blue), as defined in \cite{Christensen2015}, marked in the space spanned by the first 3 principal components of Lorenz `96. The full data set is shown in transparent black in the background.}
    \label{fig:lorenz96_regimes}
\end{figure}

To see how this topological structure relates to the more classical approach to regimes in Lorenz `96, recall the approach taken by Lorenz \cite{Lorenz2006}, further expanded on in \cite{Christensen2015}, which the reader should refer to for this discussion. In ibid, the dynamics are first projected onto the two-dimensional space spanned by the magnitudes of the concatenated principal component vectors $[PC1,PC2], [PC3,PC4]$. Two local peaks in temporal persistence are identified in this space, clearly visible in Figure 7(c) of ibid, and these are used to define two regimes denoted $A$ and $B$. Regime $A$ corresponds to the bottom right-hand corner of the concatenated space, which is also where the density is concentrated (cf. subplot (a) of the same figure), while regime $B$ corresponds to a very low-density region in the top left-hand corner. In Figure \ref{fig:lorenz96_regimes}, points loosely corresponding to these two corners of phase space have been marked, with the top left-hand corner defined by $|[PC1,PC2]|<3, 14>|[PC3,PC4]|>10$, and the bottom right-hand corner by $15>|[PC1,PC2]|>10, |[PC3,PC4]|<5$. This clearly suggests that regime $A$ corresponds to the densely populated loop around the outer perimeter, while regime $B$ corresponds to the low-density hole in the centre; we remind the reader again that the squashing away of the fourth dimension gives the appearance of regime $B$ spilling out into the perimeter. In other words, the regimes diagnosed in \cite{Christensen2015} correspond to topological features of the system that are detectable with persistent homology.

\subsection{Charney-deVore}

\begin{figure}[ht]
    \centering
    \includegraphics[scale=0.7]{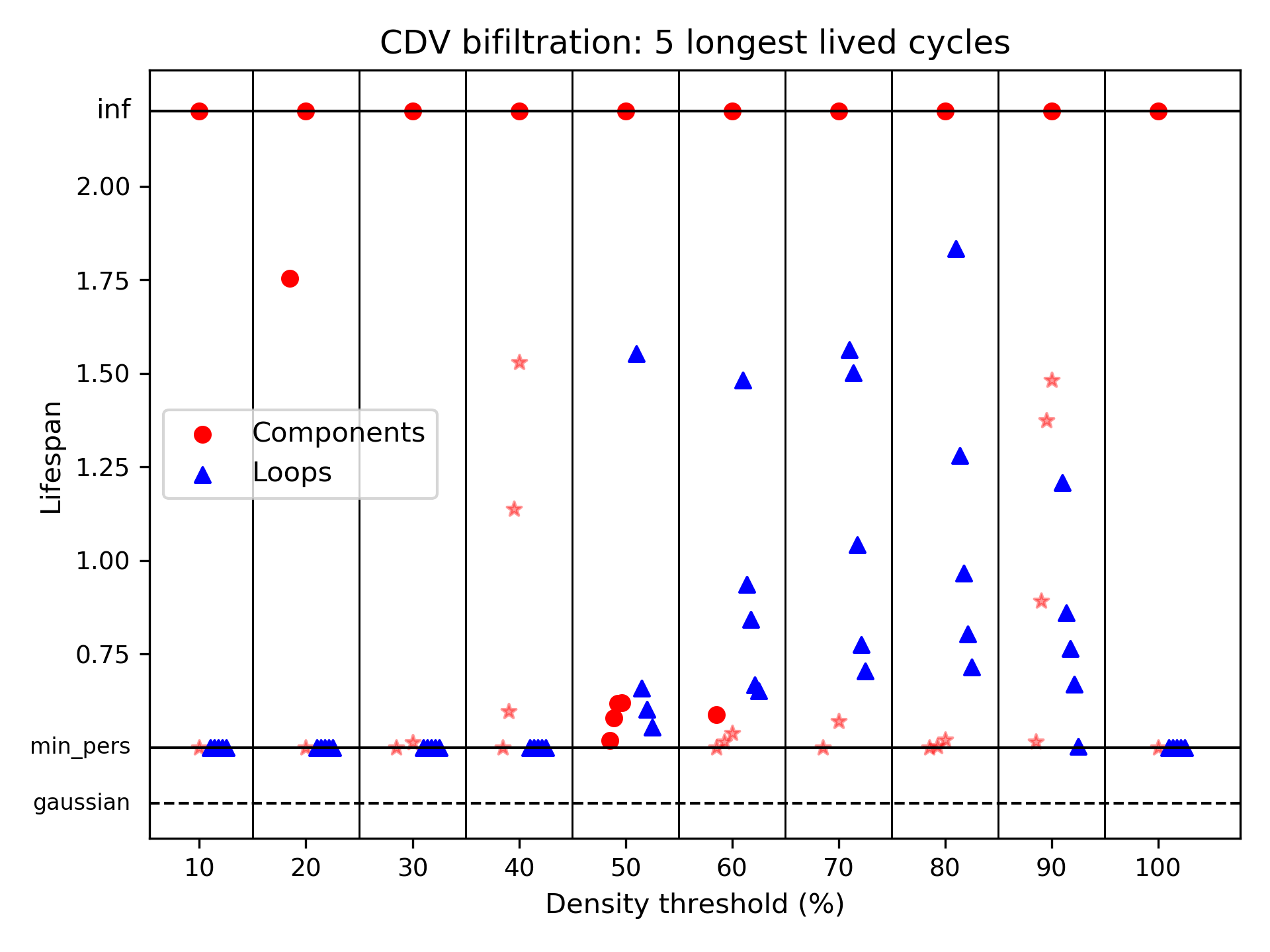}
    \cprotect\caption{A distance-density bifiltration of the CdV system. For each density threshold on the $x$-axis, the lifespan of the 5 longest-lived connected components (red dots if the component contains more than 3 points: red stars otherwise) and 5 longest-lived loops (blue triangles) are plotted. The stippled line shows the largest lifespan expected from Gaussian noise. The meaning of the \verb|min_pers| parameter is explained in Section \ref{sec:parameters}.}
    \label{fig:cdv_bifiltration}
\end{figure}

\begin{figure}[ht]
    \centering
    \includegraphics[scale=0.8]{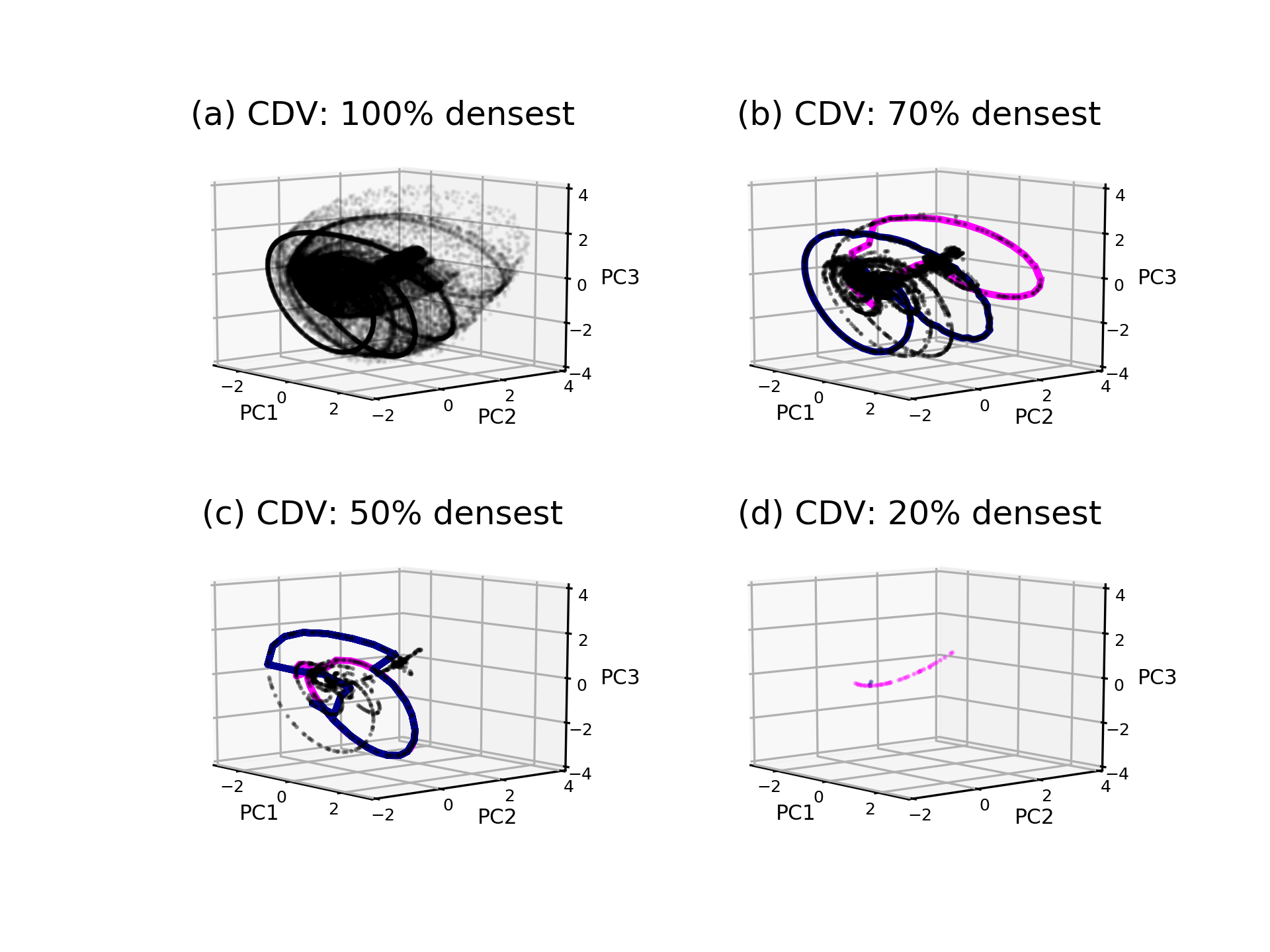}
    \caption{In (a): a long integration of the CdV system. In (b): the 70\% densest points; (c): the 50\% densest points; (d): the 20\% densest points. In (b) and (c), representatives of the 2 longest-lived loops are overlain in colour, while in (d), the 2 longest-lived components are marked in colour. In all cases, the longest-lived feature is in red and the 2nd longest-lived in blue.}
    \label{fig:cdv_thresholds}
\end{figure}

Figure \ref{fig:cdv_bifiltration} shows the bifiltration results for the CdV system: we remind the reader that the computations are done using the space spanned by the first three EOFs. The most notable features are a number of long-lived loops that emerge at density thresholds between $50\%$ and $90\%$. The existence of such loops can already be seen by eye in the raw data set, shown in Figure \ref{fig:cdv_thresholds}(a). As noted in Section \ref{sec:density_estimation}, these loops are low-dimensional, preferred trajectories shadowing unstable homoclinic orbits, separated by sparsely populated regions. The use of the direct binning method to estimate density effectively highlights these loops, and the representatives found by PersLoop, as in Figure \ref{fig:cdv_thresholds}(b) and (c), confirm that these are precisely the long-lived loops identified in Figure \ref{fig:cdv_bifiltration}.

In terms of connected components, the only threshold at which there appears to be more than one connected component with at least four points is the $20\%$ threshold. However, manual inspection here reveals that this second component in fact contains \emph{exactly} four points, and can therefore be considered as noise, as with the spurious components seen at the $40\%$ and $90\%$ thresholds. Therefore, from the perspective of the bifiltration, CdV can be thought of as a dense central region with low-density loops spiraling outward. This neatly matches the dynamics one observes in numerical simulation, and the theoretical understanding of the CdV system as chaotic transients bursting from a weakly unstable near-equilibria.

In the classical perspective, CdV has two persistent regimes associated with orbits slowing as they enter the neighbourhood of one of two fixed-points. One of these fixed points, associated with blocking, is located close to the dense central region, while the other more zonally symmetric fixed point lies close to the back left corner, when viewed as in Figure \ref{fig:cdv_thresholds}, and the loops pass close to this region. As mentioned in Section \ref{sec:cdv}, the regime dynamics in CdV are asymmetrical, in that the blocked regime is quasistationary and experiences almost deterministic evolution, while the zonal regime is characterised by turbulent chaotic behaviour. From this we can understand why the quasistationary blocking state is associated with a connected component, while the zonal state is not. Instead, the zonal regime can be understood as a consequence of the many looping trajectories visiting a common, disparate region of phase space.

\subsection{The North Atlantic Jet}

\begin{figure}[ht]
    \centering
    \includegraphics[scale=0.7]{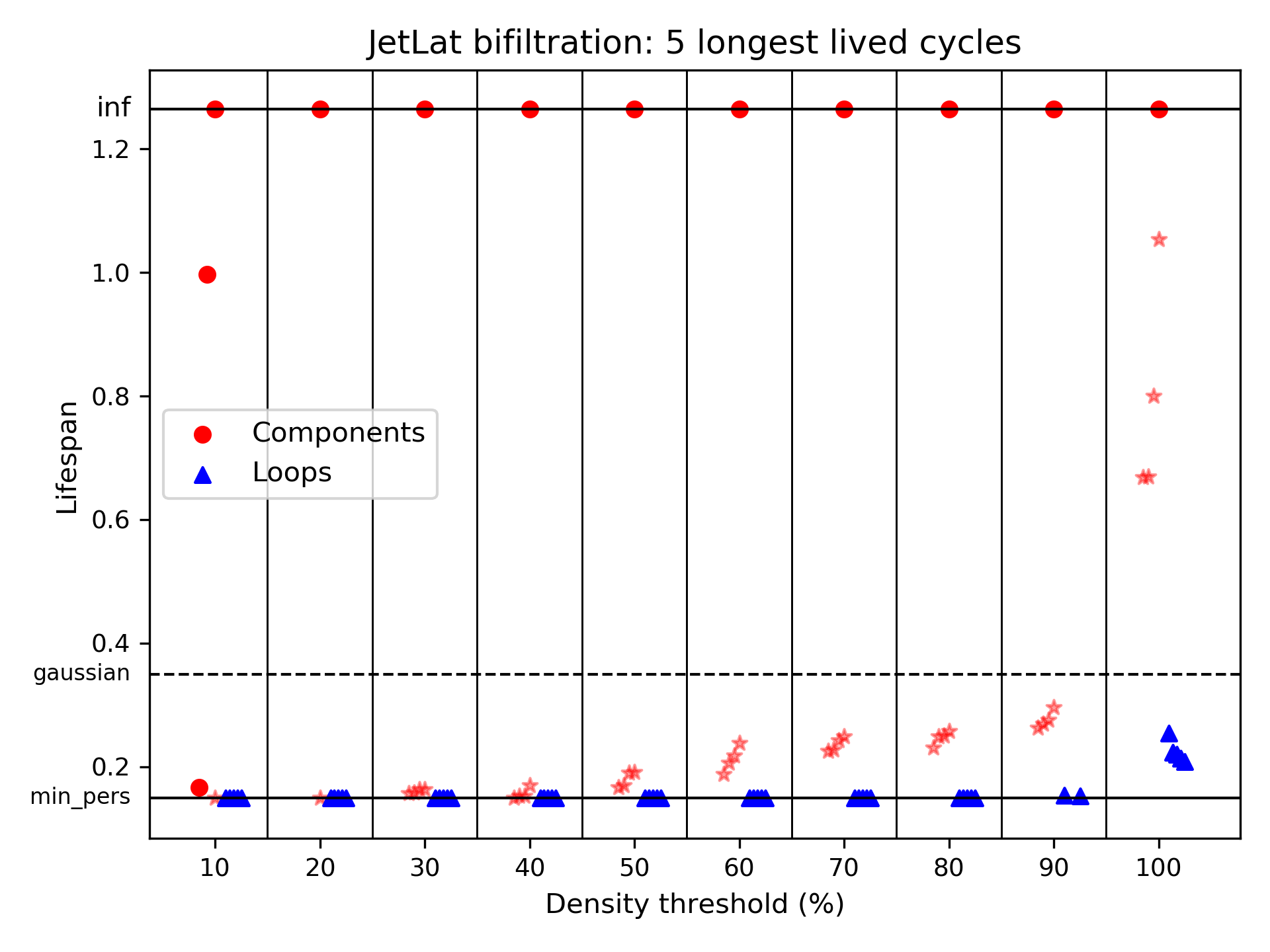}
    \cprotect\caption{A distance-density bifiltration of the JetLat data set. For each density threshold on the $x$-axis, the lifespan of the 5 longest-lived connected components (red dots if the component contains more than 3 points: red stars otherwise) and 5 longest-lived loops (blue triangles) are plotted. The stippled line shows the largest lifespan expected from Gaussian noise. The meaning of the \verb|min_pers| parameter is explained in Section \ref{sec:parameters}.}
    \label{fig:jetlat_bifiltration}
\end{figure}

\begin{figure}[ht]
    \centering
    \includegraphics[scale=0.8]{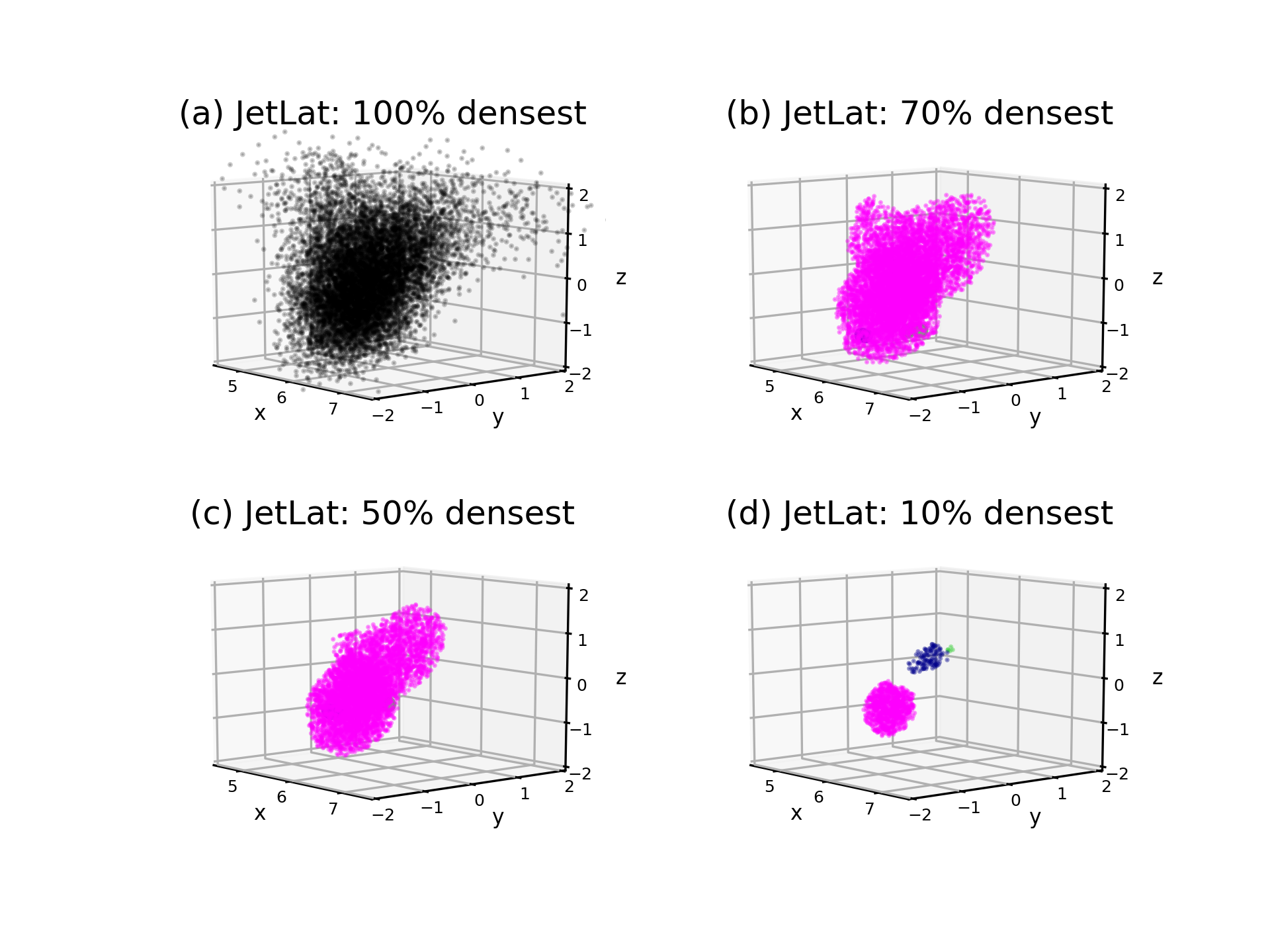}
    \caption{In (a): the raw JetLat data set. In (b): the 70\% densest points; (c): the 50\% densest points; (d): the 10\% densest points. In (b)-(d), representatives of the 3 longest-lived connected components are marked with colours. The longest-lived feature is in red, the 2nd longest-lived in blue and the 3rd longest-lived in green.}
    \label{fig:jetlat_thresholds}
\end{figure}

\begin{figure}[ht]
    \centering
    \includegraphics[scale=0.7]{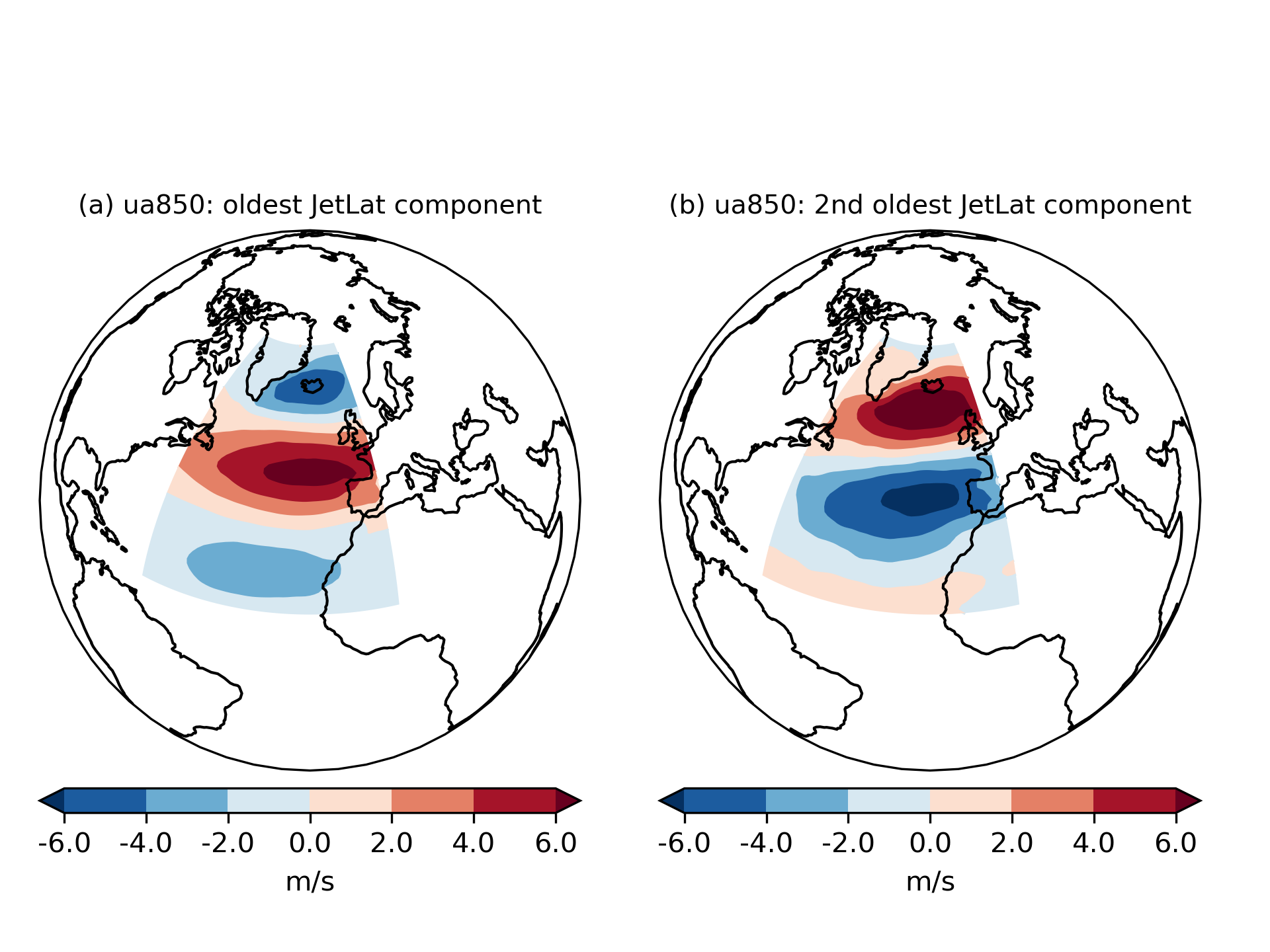}
    \caption{Composites of zonal wind anomalies at 850hPa for the ERA20C data set across all winter days between 1900 and 2010 that (a) belong to the longest-lived JetLat component; (b) belong to the 2nd longest-lived JetLat component.}
    \label{fig:jetlat_composites}
\end{figure}

We finally test our method using the JetLat data set, capturing variability of the North Atlantic eddy-driven jet. Figure \ref{fig:jetlat_bifiltration} shows the result of the bifiltration computation, with Figure \ref{fig:jetlat_thresholds} visualising select thresholds. The only evidence of non-trivial topology emerges when restricting to the 10\% densest points, at which point the data set splits cleanly into two connected components, as shown in Figure \ref{fig:jetlat_thresholds}(d). The lifespans of both components greatly exceed anything expected from Gaussian noise, and their sizes are also considerable, containing around 900 and 100 points each. Figure \ref{fig:jetlat_composites} shows composites of zonal wind anomalies of ERA20C across all days belonging to these two long-lived components, identifying the longest-lived one as the Central jet latitude mode and the 2nd longest-lived as the Northern jet latitude mode. 

\begin{figure}[ht]
    \centering
    \includegraphics[scale=0.75]{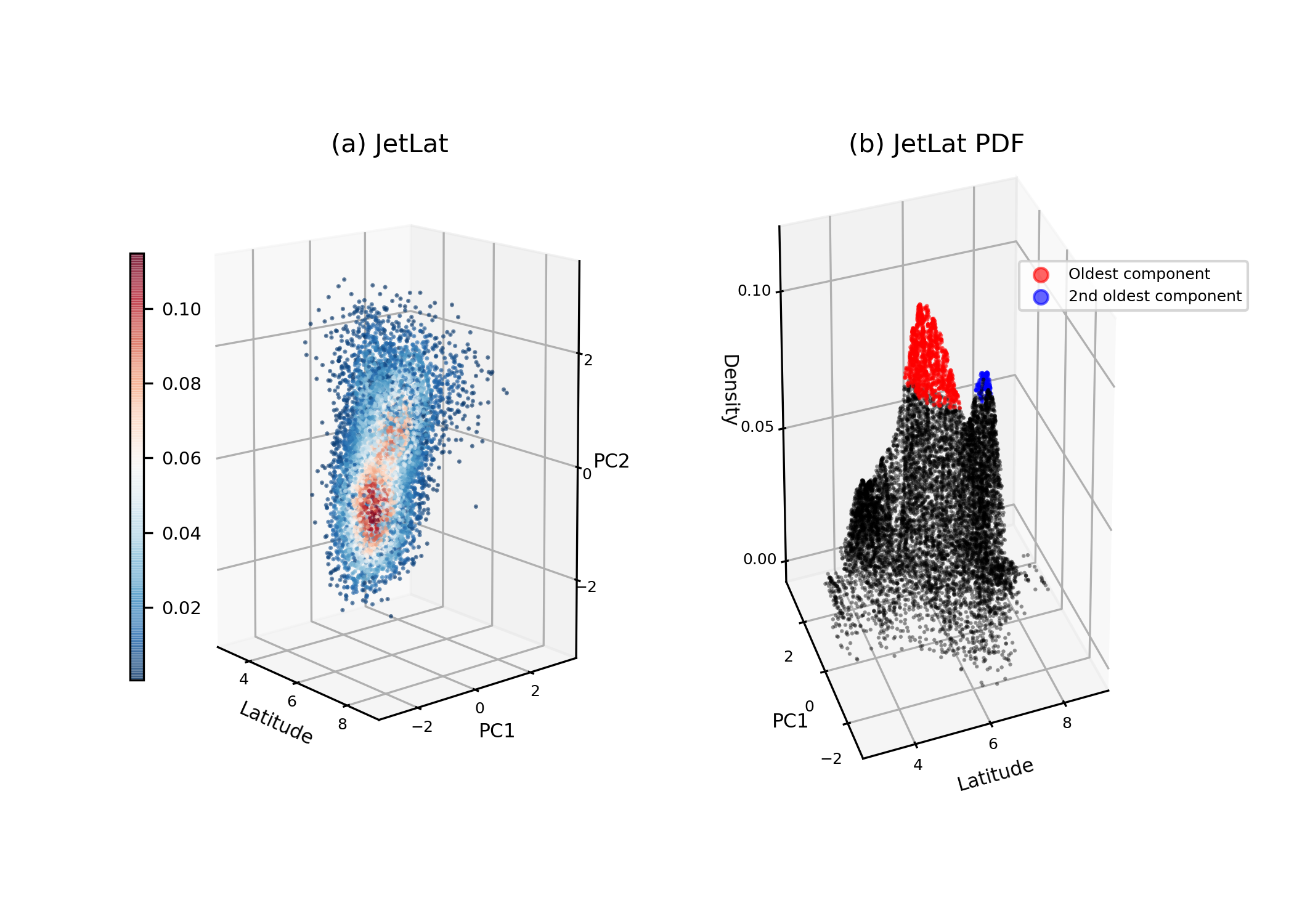}
    \caption{Visualising the pdf of the JetLat data set. In (a), the 3-dimensional data with colours indicating the density, as estimated with a kernel density estimator. In (b), density as a function of the two first dimensions of the JetLat data set, i.e. jet latitude and the first principal component of ua850. In (b), points belonging to the longest-lived connected component are coloured red, while points belonging to the 2nd longest-lived component are coloured blue.}
    \label{fig:jetlat_pdf}
\end{figure}

Since one dimension of the JetLat data set contains the jet latitude index, which is trimodal in and of itself, the a priori expectation might be that the data set should split into three connected components, not two. However, making the density filtration finer did not change the result, suggesting this is a robust outcome of our methodology. To understand why this happens, Figure \ref{fig:jetlat_pdf} shows the JetLat probability distribution function (pdf), as computed using the kernel density estimator. In panel (a), the raw data set is plotted with colours indicating density, while in (b), density is plotted as a function of jet latitude and $PC1$ (the first two dimensions of JetLat). In this latter panel, the points corresponding to the two long-lived components at the $10\%$ threshold have been coloured in, with red being the longest-lived and blue the 2nd longest-lived. While panel (a) already suggests that there are two, rather than three, clearly marked peaks in density, panel (b) most clearly explains what is happening. In and of itself, the jet latitude index is clearly trimodal, but the situation changes when it is extended out across multiple dimensions. While the Northern peak remains clearly separated from the Central peak, the Southern peak becomes smeared out across the space spanned by the two principal components, leaving it resembling a `shoulder', rather than a clear peak. Because our density thresholds amount to taking horizontal slices across this space, the bifiltration is able to find the Central and Northern peaks, but not the Southern. The implications of this are discussed in the next section.

Note that when computing a bifiltration using the first 3, 4 or 10 principal components of geopotential height anomalies at 500hPa, the features detected are all at the level of Gaussian noise. This is consistent with the findings of \cite{Stephenson04}.

\section{Discussion}
\label{sec:discussion}

\subsection{Strengths and weaknesses of our methodology}

The results in the previous section suggest that the bifiltration methodology succeeds in identifying whether a data set has non-trivial topological structure or not. In particular, it rejects a Gaussian distribution as having any, and correctly detects the relevant structure for four examples of data sets generally considered to have regimes. We further showed that the topological structure encodes, in different ways, the regime behaviour. For Lorenz `63, the regimes correspond to two holes; for Lorenz `96 to a loop and a hole; for CdV to a dense, connected region and several loops emanating from this; and for JetLat, to two dense, connected components. The four systems considered are thereby clearly distinguished through their differing homology groups.

The main apparent shortcoming of the methodology, besides the instabilities associated with trying to compute optimal representatives of loops, was the inability to identify three distinct regimes in the JetLat data set. As explained in Section \ref{sec:results}.5, this is due to the fact that, when viewed across multiple dimensions, the Southern jet latitude mode appears less as a distinct peak and more as an extended shoulder, which the horizontal density slices of our filtration cannot easily capture. The obvious way to attempt to remedy this is to consider slices with positive slope. As explained in Section \ref{sec:bifiltrations}, this is also required to make the evolution of topological features continuous across the bifiltration, implying that this is a natural way to improve our methodology for stability reasons alone. We hope to examine this, using the RIVET software (cf. Section \ref{sec:rivet}), in future work. It has also been noted \cite{Hazelton2003} that Gaussian kernels can sometimes flatten peaks too much: a more thorough examination of optimal density estimators for our data sets is for this reason another avenue of future work.

While the failure to detect the Southern jet mode should probably be viewed as a shortcoming, we would also suggest that this failure may shed some light on a few curious features in the literature. Firstly, many studies have tried to diagnose regimes in the Euro-Atlantic sector, and, depending on the choice of input data, pre-processing steps and diagnostics, these studies have suggested there may be anywhere between 2 and 6 regimes (see \cite{Hannachi2019}, \cite{Dorrington2020}, \cite{Dawson2012}, \cite{Madonna2017} and \cite{Falkena2020} respectively for examples of each number). While the ambiguity between the choices 3, 4 and 5 is at least in part due to the confounding influence of the jet speed \cite{Dorrington2020}, and the choice of 2 regimes usually corresponds to the North Atlantic Oscillation dipole \cite{Woollings2008, Hannachi2019}, the striking divergence in the number of regimes across studies using similar techniques is still somewhat puzzling. Our results suggest that one possible reason for this is that, depending on what angle one views the Euro-Atlantic circulation from, different regimes may appear either as clearly distinct peaks or more ambiguous and hard to detect shoulders.

Secondly, in \cite{Strommen2020}, the ability of a numerical weather forecast model to make skillful predictions of the Euro-Atlantic circulation was studied from the perspective of the three jet latitude regimes. It was found that the model was able to skillfully detect changes in the Northern mode compared to the Southern and Central modes, but was not able to robustly separate between the Southern and Central modes. In other words, from the perspective of the forecast model, the jet appeared to behave as if it had 2, not 3, regimes. By considering Figure \ref{fig:jetlat_pdf}(b), it is perhaps not surprising that an imperfect model may struggle to reproduce the more subtle behaviour of the Southern shoulder, and produce a cruder approximation of the pdf as having just two peaks. A comparison between this figure and an equivalent one for model data (not shown) does suggest the model has a notably flatter Southern peak.

\subsection{Why a simpler definition of regime fails}

We have shown that non-trivial topological structure, as measured with a bifiltration of homology, provides a unifying way of understanding the main examples of non-linear dynamical systems generally considered as exemplifying regime behaviour. Because this comes at the cost of introducing an extra level of abstraction, it is reasonable to ask if a similar unification could be achieved using the more common ways of understanding regimes, namely density peaks (i.e., clustering) or temporal persistence. We will now show that, on the face of it, no such simpler unification appears possible.

To see this, first notice that while for JetLat, the two regimes correspond clearly to local maxima in density, both Lorenz `96 and CdV are examples where the density of the two regimes are wildly different. For Lorenz `63, while a bimodal pdf can be obtained by time-averaging \cite{Corti1999}, Figure \ref{fig:lorenz_holes_evolution}(d) makes it clear that the regions defined by the two regimes (i.e., the two wings) are, in the raw data set, not local density maxima. Hence a definition of regimes as local density maxima/minima or clustering will invariably fail to account for one of these systems. 

Next, one might consider a criteria based on any of the closely related concepts of temporal persistence, average residence times or phase speed velocities. However, also here one finds that the behaviour of the different systems differs dramatically. In the Lorenz `63 system, temporal persistence peaks (and velocities are smallest) at the dense region connecting the two wings, while temporal persistence is in general minimal in the wings themselves, where velocities peak; the exception being the extremely rare trajectories that pass sufficiently close to either fixed point. On the other hand, for Lorenz `96, both regimes correspond to peaks in temporal persistence/residence time, as mentioned already, while in CdV the two regimes are broadly asymmetric in terms of their temporal persistence and velocities, with the blocking regime featuring high temporal persistence/low velocities and the zonal regime favouring low temporal persistence/high velocities. Even in the real atmosphere, the behaviour does not appear to be uniform. Already in \cite{Woollings2010b}, where the jet latitude regimes were first presented, it was noted that the forcing on the jet by transient eddies, thought to be a key driver in generating temporal persistence, appears to be operating similarly at all latitudes, not just at the peaks of the trimodal distribution. In other words, the extent to which the three jet latitude modes can be characterised as having higher-than-average temporal persistence is ambiguous. This ambiguity is further supported by the results of \cite{Faranda2017}, which examined the closely related 4-regime picture of the North Atlantic. By computing a measure of both local temporal persistence and local density, they locate the four regimes in distinct quadrants of temporal persistence-density space, implying the regimes all have strikingly different characteristics.

A simplistic definition of regimes based on temporal persistence, residence times or velocities will, therefore, inevitably fail to capture the behaviour in one or more of these systems. It is also clear from this discussion that the situation cannot be salvaged by using a definition combining both notions. Hence it seems, to these authors, not to be possible to find a definition of regimes, using density or temporal persistence alone, that unifies all the systems we considered. While an alternative definition of regimes based on fixed points, UPOs or other `exact solution' techniques might seem plausible, computing such solutions is extremely computationally demanding, and state-of-the-art techniques are only able to handle systems of significantly lower dimensionality than existing climate models \cite{lucarini2020new}. More crucially, these techniques are inherently model features, in that they rely on being able to integrate the model dynamics. Given that models are known to exhibit systematic biases in their regime structure \cite{Fabiano2020}, inferring conclusions about the real atmosphere based on results obtained from models would require considerable care. It is therefore not currently clear how such `exact solution' techniques can be applied to observational data sets.

Some readers may reasonably question whether it is in fact important to have a unified framework for understanding regimes, and that the word `regime' is perhaps best understood as a context-dependent phrase that captures a wide variety of ways to simplify complex, non-linear dynamics. Indeed, it is possible that there exist dynamical systems that appear to exhibit regime behaviour that cannot be accounted for by topological means. Nevertheless, the fact that four very different systems do allow for such a topological characterisation lends confidence to this being possible in a wide variety of cases. Furthermore, we believe that the more ad hoc regime approach common in atmospheric science - and the lack of any clear unifying framework - has in general undermined confidence both in their practical usage and even their existence \cite{Stephenson04, Christiansen2007, Fereday2017}. The existence of non-trivial topological structure underpinning four quintessential examples found in the literature may help bolster confidence that the various attempts to diagnose regimes in the atmosphere are really characterising genuine features of the climate attractor.

\section{Conclusions and further directions}
\label{sec:conclusions}

In this paper we have argued that the unifying feature across the most well-known examples of regime systems is their non-trivial topological structure. We showed that, using persistent homology, one can compute a bifiltration of topological invariants which encodes such non-trivial structure. By carrying out this computation for four classical regime systems (Lorenz `63, Lorenz `96, Charney-deVore and the North Atlantic jet), we showed that the information detected in such a bifiltration encodes the key features of each system associated with their regimes. It was pointed out that these systems also exhibit widely differing behaviour in terms of the density and temporal persistence of their regimes, suggesting that no simple definition of regime structure based solely on these notions is likely to be general enough to capture all of them.

These results justify our suggestion that the notion of a regime in a dynamical system can be understood as the results of varied attempts to capture the non-trivial topology of the underlying attractor. This approach can be obviously adjusted to relate to local regions of phase space only, to account for, e.g., the Euro-Atlantic sector as a particular region in the larger climate attractor. Our methodology shows that besides being an approach which captures a sufficiently wide variety of behaviour, it has the important quality of being computationally tractable for the size of data sets typically used in meteorology and climate science. Furthermore, far from being simply a mathematically neat abstraction, we argue that this topological perspective on regimes offers concrete practical benefits, for three main reasons.

To understand the first reason, it is helpful to recall, as discussed in the introduction, that the raison d'être of regimes is to understand questions of predictability across multiple timescales. An overemphasis on properties related to density (as in clustering methods) or temporal persistence may end up obfuscating analysis, not only because regime systems can have a wide variety of behaviour with respect to these notions, but, crucially, because the most salient information may be located in entirely different aspects of the system. The CdV system is an instructive example in this regard. While its classical regimes are associated with fixed points, the most striking impact of these is the tight, looping behaviour it generates (cf. Figure \ref{fig:cdv_thresholds}). Knowing that the system is on such a narrowly defined trajectory provides significantly more information than simply knowing that the system is in the vicinity of a fixed-point. From a topological perspective, where no knowledge of fixed-points is implicit, these loops are what stand out as the major feature of CdV, implying that focusing attention on such features can highlight information which is otherwise being overlooked. This potential of topological methods to obtain efficient, simplified representations of chaotic dynamics was also noted in \cite{Gokhan2020} using different ideas.

The second and third reasons relate to the technical benefits of persistent homology algorithms. Unlike many existing algorithms for regime analysis, such as $K$-means clustering, persistent homology is effectively non-prescriptive. That is, the only parameters required for the algorithm are generic to the system, such as a measure of the spatial scales of the system, as opposed to parameters that explicitly influence the diagnosed regimes, such as the choice of $K$ in $K$-means. Homological techniques are therefore particularly well-suited to studying systems where prior knowledge of regime structure is less clear. The ability of our technique to capture the regime behaviour associated to several classical systems lends confidence in its ability to locate relevant structure in such contexts, such as the real atmosphere.

Finally, we note that several studies suggest that the multimodal behaviour observed in the North Atlantic jet involves not just changes to zonal winds, but the complex interplay between winds, pressure and temperatures, in the form of meridional heat transport and baroclinicity \cite{Novak2015}. There are tantalising clues that genuine loops in the climate attractor might be detectable when taking this into account (cf. \cite{Novak2017}, Figures 4 and 5). It is therefore plausible that the detection of such loops in the climate attractor requires a technique that can gracefully handle multiple dimensions of data encoding several atmospheric variables. The fact that homological computations are effectively exempt from the `curse of dimensionality' (cf. Section \ref{sec:complexity}) implies that persistent homology is, in principle, such a tool. Turning this principle into realisable practice will, however, require improvements to two branches of software: software capable of carrying out more subtle bifiltrations (such as RIVET), and software capable of producing stable optimal representatives of homology classes (such as PersLoop). It is the hope of these authors that such improvements might allow for the detection of robust regime structure in the atmosphere using unprocessed, but high-dimensional, observational data.

\section*{Acknowledgements}
We thank Sayan Mandal and Tamal Dey for their assistance with the software package PersLoop, as well as Hannah Christensen for sharing Lorenz `96 data and helpful conversations. We thank Mason A. Porter for helpful feedback.

KS was funded by a Thomas Philips and Jocelyn Keene Junior Research Fellowship in Climate Science at Jesus College, Oxford. JD is funded by NERC Grant NE/L002612/1. MC was supported by a grant from the Office of Naval Research Global. 

\section*{Data Availability and Code}

Python code supporting the computations of this paper, along with the data used in this study, can be found at \url{https://github.com/KristianJS/BifiltPH}.

\begin{appendices}

\section{Persistent homology}\label{SS:definition of barcode}

In this section we provide the definition of barcodes for the persistent homology of a finite metric space with respect to the Vietoris--Rips complex.   Most of the material in this section only requires a linear algebra background (with the exception of Proposition \ref{P:decomposition} for which more advanced algebraic notions are needed). We  refer the reader 
 to \cite[Section 4.1]{otter2017roadmap} and references therein for alternative definitions of barcodes, as well as further details and intuition on the concepts presented here. 

\begin{definition}
A {\bf simplicial complex} $K=(V,\Sigma)$ is given by a set $V$ together with a collection $\Sigma$ of subsets of $V$ satisfying that (i) $\{v\}\in \Sigma$ for all $v\in V$ and (ii) if $\sigma\in \Sigma$ and $\tau\subset \sigma$, then $\tau\in \Sigma$. We call the elements of $V$ the {\bf vertices} of the simplicial complex, while the elements of $\Sigma$ are called {\bf simplices}. A {\bf $p$-simplex} is a simplex with cardinality $p-1$, and we say that $p$ is the {\bf dimension} of such a simplex. 
\end{definition}

We note that what we call ``simplicial complex'' is usually called  ``abstract simplicial complex'' in the literature. Every simplicial complex with a finite set of vertices can be realised as a subset of Euclidean space in a canonical way, by identifying the vertices with the standard basic unit vectors, and  one can intuitively think of a simplicial complex as a subspace of Euclidean space obtained by gluing together vertices, edges and higher dimensional simplices along their common faces. In what follows we encourage the reader to keep this intuition in mind, and to think of $0$-simplices as points in Euclidean space, $1$-simplices as closed straight line segments, $2$-simplices as triangle-shaped closed convex subspaces, and so on.

\begin{example}
Consider  the simplicial complex with set of vertices $V=\{a,b,c\}$ and set of simplices $\Sigma=\{\{a\},\{b\},\{c\},\{a,b\},\{b,c\}\}$. We can realise $(V,\Sigma)$ as a subset of $\mathbb{R}^3$ by identifying the vertex $a$ with the  vector $(1,0,0)$, the vertex $b$ with the  vector $(0,1,0)$ and the vertex $c$ with the  vector $(0,0,1)$. The $1$-simplex $\{a,b\}$ is then identified with the straight line segment connecting the points $(1,0,0)$ and $(0,1,0)$, while the $1$-simplex $\{b,c\}$ is identified with the straight line segment connecting the points $(0,1,0)$ and $(0,0,1)$.
\end{example}

\begin{definition}
Let $(X,d)$ be a finite metric space, and let $\epsilon$ be a non-negative real number. The { \bf Vietoris--Rips complex at scale $\epsilon$} is the simplicial complex $V(X)(\epsilon)$ whose set of vertices is given by $X$, and such that $\sigma\subset X$ is in  $V(X)(\epsilon)$ if and only if $d(x_i,x_j)\leq \epsilon$ for all $x_i,x_j\in \sigma$.  
\end{definition}

\begin{example}\label{E:VR}
Consider $X=\{x_0,x_1,x_2,x_3\}$ with the following distances:\\
$
d(x_0,x_1)=1,\; d(x_0,x_2)=1,  \;d(x_0,x_3)=1.2, \; d(x_1,x_2)=1.1, \; d(x_1,x_3)=0.5 $
 and $d(x_2,x_3)=0.6$.
We then have that $V(X)(0.1)$ is a simplicial complex with the four $0$-simplices $\{x_0\},\{x_1\}$, $\{x_2\}$ and $\{x_3\}$, and with no higher dimensional simplices. On the other hand, $V(X)(1)$  is a simplicial complex  with the four $0$-simplices $\{x_0\},\{x_1\}$, $\{x_2\}$ and $\{x_3\}$ and the four $1$-simplices given by $\{x_0,x_1\}$, $\{x_1,x_3\}$, $\{x_0,x_2\}$ and $\{x_2,x_3\}$. Further, $V(X)(1.2)$ is the simplicial complex with the four $0$-simplices $\{x_0\},\{x_1\}$, $\{x_2\}$ and $\{x_3\}$, six $1$-simplices  $\{x_0,x_1\}$,  $\{x_0,x_2\}$, $\{x_0,x_3\}$, $\{x_1,x_2\}$,  $\{x_1,x_3\}$ and $\{x_2,x_3\}$, four $2$-simplices $\{x_0,x_1,x_2\}$, $\{x_1,x_2,x_3\}$, $\{x_0,x_1,x_3\}$,$\{x_0,x_2,x_3\}$ and one $3$-simplex $\{x_0,x_1,x_2,x_3\}$. \end{example}
As we see in Example \ref{E:VR}, as we increase the scale value we are adding more and more simplices to the simplicial complex. 
In general, we have that if $\epsilon\leq \epsilon'$ then the set of simplices of  $V(X)(\epsilon)$ is contained in that of  $V(X)(\epsilon')$.

\begin{definition}
Let $\mathbb{F}_2$ be the field  with two elements, and $K=(V,\Sigma)$ a simplicial complex. For each $p=0,1,2,\dots $ we define $C_p$ to be the $\mathbb{F}_2$-vector space with basis given by the $p$-simplices of $K$. Furthermore, we define linear maps $d_{p+1}\colon C_{p+1}\to C_{p}$ for all $p=0,1,2,\dots$ as follows: 
\[
d_{p+1}(\sigma)=\underset{\text{s.t.\,}\tau\subset \sigma }{\sum_{\tau\in \Sigma_p}}\tau \, ,
\]
where $\sigma$ is any $p+1$-simplex in $K$. We set $d_0:C_0\to 0$ to be the map sending every element of $C_0$ to $0$, where $0$ denotes the zero vector space.
The collection of vector spaces $C_p$ and linear maps $d_p$ for $p=0,1,2,\dots$ is called the {\bf simplicial chain complex} of $K$.
\end{definition}

\begin{example}\label{E:CC VR}
We consider  the Vietoris--Rips simplicial complex  $V(X)(1)$ from Example  \ref{E:VR}. We have that $C_0$ is a vector space of dimension $4$ with basis given by the four $0$-simplices $\{x_0\}$, $\{x_1\} $, $\{x_2\}$ and $\{x_3\}$, while  $C_1$ is a vector space of dimension $4$ with basis given by the four $1$-simplices $\{x_0,x_1\}$, $\{x_1,x_3\}$, $\{x_0,x_2\}$ and $\{x_2,x_3\}$. Since there are no simplices of dimension $2$ or higher, $C_p$ is the trivial vector space for $p\geq 2$.  The map $d_1\colon C_1\to C_0$ is defined as follows: we have $d_1(\{x_0,x_1\})=\{x_0\}+\{x_1\}$,  $d_1(\{x_1,x_3\})=\{x_1\}+\{x_3\}$, and similarly  $d_1(\{x_0,x_2\})=\{x_0\}+\{x_2\}$, and $d_1(\{x_2,x_3\})=\{x_2\}+\{x_3\}$. Keeping the geometric intuition in mind, we can think of $d_1$ as sending each closed straight-line segment to the sum of the two points in its boundary. Furthermore, we have that $d_p=0$ for any $p\ne 1$.

\end{example}

As one can easily compute, we have that $d_p\circ d_{p+1}=0$ for all $p$, and therefore  the image of the map $d_{p+1}$ is contained in the kernel of the map $d_{p}$.  The elements of the image of $d_{p+1}$ are called {\em $p$-boundaries}, while the elements of the kernel of $d_p$ are called {\em p-cycles}. We note that one can more generally define a simplicial chain complex for any field, but one would need to take care of defining the maps $d_p$ differently, to ensure that the composition of any two consecutive maps yields the zero map. In the computations that we perform in this manuscript we use  the field with two elements.

\begin{definition}\label{D:simplicial homology}
Let  $p$ be a natural number. The {\bf $p$th simplicial homology} of a simplicial complex $K$ is the $\mathbb{F}_2$-vector space $H_p(K)=\mathrm{ker}(d_p)/\mathrm{im}(d_{p+1})$. One calls the rank of $H_p(K)$ the {\bf $p$th Betti number} of $K$, or alternatively, the number of  {\bf $p$-dimensional holes} of $K$. 
\end{definition}

The $p$th Betti number thus measures the number of $p$-cycles that are not $p$-boundaries. Intuitively, for $p=1$, a cycle which is not a boundary corresponds to a loop which doesn't contain anything in its interior, in other words a hole. For instance, for the Vietoris--Rips complex $V(X)(1)$ from Example \ref{E:CC VR}, we have that the $1$st Betti number is one: we have that $\{x_0,x_1\}+ \{x_1,x_3\}+\{x_0,x_2\}+\{x_2,x_3\}$ is in the kernel of $d_1$ and thus is a $1$-cycle, however, it is not in the image of $d_2$, since there are no simplices of dimension $2$ or higher. On the other hand, if we consider the Vietoris--Rips complex $V(X)(1.1)$, then we have that its $1$st Betti number is zero. Here we again have that $\{x_0,x_1\}+ \{x_1,x_3\}+\{x_0,x_2\}+\{x_2,x_3\}$ is in the kernel of $d_1$, but now it is also   the $1$-boundary of $\{x_0,x_1,x_2\}+\{x_1,x_2,x_3\}$. We discuss some of these computations in detail in the following example.

\begin{example}\label{E:simpl hom}
We compute  $p$th simplicial homology for the simplicial complex in Example \ref{E:CC VR}. For $p=0$, we have that $H_0(V(X)(1))=C_0/\mathrm{im}(d_{1})$. The map $d_1$ has a $3$-dimensional image spanned by  $\{x_0\}$, $\{x_1\}+\{x_3\}$ and $\{x_0\}+\{x_1\}+\{x_2\}$. Thus, the quotient $C_0/\mathrm{im}(d_{1})$ is $1$-dimensional. 

For $p=1$, we have that $H_1(V(X)(1))=\mathrm{ker}(d_1)/\mathrm{im}(d_{2})\cong\mathrm{ker}(d_1)$, since $d_2$ is the zero map. 
The kernel of $d_1$ is $1$-dimensional, and is spanned by $\{x_0,x_1\}+\{x_1,x_3\}+\{x_0,x_2\}+\{x_2,x_3\}$. Furthermore, we have that the $p$th simplicial homology vector space is zero-dimensional for $p\geq 2$. Thus, we have that the simplicial complex $V(X)(1)$ has one $0$-hole,  one $1$-hole, and no higher dimensional holes. A choice of basis vector for  $H_0(V(X)(1))$ is given by $[\{x_0\}+\{x_1\}+\{x_2\}+\{x_3\}]$, while a generator for $H_1(V(X)(1))$ is $[\{x_0,x_1\}+\{x_1,x_3\}+\{x_0,x_2\}+\{x_2,x_3\}]$, where the square brackets denote equivalence classes in the respective quotient vector spaces. Intuitively, we can think of the generator of $H_1(V(X)(1))$ as consisting of the ``loop'' formed by gluing together the four straight line segments corresponding to the $1$-simplices along their common boundary points.
\end{example}

Any map of simplicial complexes $f\colon K\to K'$, namely a map between the sets of vertices sending simplices to simplices, induces a linear map $H_p(f)\colon H_p(K)\to H_i(K')$ between the respective homology vector spaces. Here we are interested in inclusion maps between Vietoris--Rips complexes associated to a metric space $X$, at different scale values. 

By computing simplicial homology of $V(X)(\epsilon)$  for any real number $\epsilon\geq0$, we obtain what is called a  persistence module:
\begin{definition}
A {\bf persistence module} is a collection of  $\mathbb{F}_2$-vector spaces $\{M_\epsilon\}_{\epsilon\in \mathbb{R}_{\geq 0}}$ together with a collection of  $\mathbb{F}_2$-linear maps $\phi(\epsilon,\epsilon')\colon M_\epsilon\to M_{\epsilon'}$ for all $\epsilon\leq \epsilon'$ in $\mathbb{R}_{\geq 0}$ such that 
\[
\phi(\epsilon',\epsilon'')\circ \phi(\epsilon,\epsilon')=\phi(\epsilon,\epsilon'')
\]
whenever $\epsilon\leq \epsilon'\leq \epsilon''$. The linear maps $\phi(\epsilon,\epsilon')$ are called {\bf structure morphisms}.

\end{definition}

Computing $p$th simplicial homology of the Vietoris--Rips complexes  at all scales, we obtain a persistence module given by $\{H_p(V(X)\epsilon)\}_{\epsilon\geq 0 }$, together with the structure morphisms  $H_p(\iota_{\epsilon,\epsilon'})\colon H_p(V(X)(\epsilon))\to H_p(V(X)(\epsilon'))$ for all $\epsilon\leq \epsilon'$ which are  given by the linear maps induced by the inclusions $\iota_{\epsilon,\epsilon'}\colon V(X)(\epsilon)\to V(X)(\epsilon')$. 
While the $p$-th betti numbers capture information about the number of $p$-dimensional holes at each scale, and for instance the number of connected components, or loops, for $p=0$ or $1$, respectively, $p$th persistent homology also captures  information about how the connected components or loops change as we increase the scale parameter.
More precisely, one has that such a persistence module satisfies a finiteness condition, which ensures that the following holds:

\begin{proposition}\label{P:decomposition}
Let $(X,d)$ be a finite metric space, and let $V(X)(\epsilon)$ be the Vietoris--Rips complex of $V$ at scale $\epsilon$. The persistence module $(\{H_p(V(X)(\epsilon))\}_{\epsilon \geq 0}\},\{H_p(\iota_{\epsilon, \epsilon'})\}_{\epsilon\leq \epsilon'})$ is  finitely generated, and  we have that there  exists an $m\in \mathbb{N}$ and  unique (up to reindexing) intervals $[b_i,d_i)$ for $i=1,\dots , m$ such that 
\begin{equation}\label{E:decomposition}
\bigoplus_{\epsilon\geq 0}{H_p(V(X)(\epsilon))}\cong \bigoplus_{i=1}^m \mathbb{I}({[b_i,d_i}))\, .
\end{equation}
\noindent
Here, for non-negative reals $a< b$, we denote by   $\mathbb{I}({[a,b}))$  an interval module, namely a persistence module with 
\[
\mathbb{I}({[a,b}))(\epsilon)=
\begin{cases}
\mathbb{F}_2, & \epsilon\in [a,b)\\
0, &\text{ otherwise}
\end{cases}
\]
and the structure morphisms $\phi_{\epsilon,\epsilon'}$ are given by the identiy linear map whenever ${a\leq \epsilon\leq \epsilon'<b }$, and the zero map otherwise.

\end{proposition}

\begin{remark}
We note that   in Eq.~\ref{E:decomposition}  the direct sums of vector spaces are endowed with additional structure that comes from identifications encoded by the structure morphisms: these direct sums are  graded modules over certain monoid rings, and the finiteness condition is understood as being a condition on these modules. We refer the reader to \cite{CK18} and references therein for more details.
\end{remark}

We can understand the decomposition in  Eq.~\ref{E:decomposition}  as follows: there exists a choice of basis vectors at each scale value $\epsilon$,  such that we can represent the information given by the homology vector spaces and linear maps between them in a diagram consisting of disjoint intervals, called a {\em barcode}.  For such a choice of basis vectors, we say that $x\in H_p(V(X)(\epsilon))$ is {\em born at $\epsilon$} if it is not in the image of $H_p(\iota_{\epsilon',\epsilon})$ for any $\epsilon'<\epsilon$. Similarly, we say that  $0\ne x\in H_p(V(X)(\epsilon))$ {\em dies at $\epsilon''$}, for $\epsilon''>\epsilon$ if $\epsilon''$ is the smallest scale value so that $H_p(\iota_{\epsilon,\epsilon''})(x)=0$.

In Figure \ref{F:example PH} in the main text we provide an example of  barcode: in purple we depict the barcode plot  the persistence module  $(\{H_1(V(X)(\epsilon))\}_{\epsilon\geq 0}, \{H_1(\iota_{\epsilon,\epsilon'})\}_{\epsilon\leq \epsilon'})$, and in blue the barcode  for the persistence module ${(\{H_0(V(X)(\epsilon))\}_{\epsilon\geq 0},\{H_0(\iota_{\epsilon,\epsilon'})}_{\epsilon\leq \epsilon'})$, where $X$ is the finite set of points in Figure  \ref{F:example PH}(a). 

\begin{example}\label{E:barcode}
We consider again our running example, namely the Vietoris--Rips complexes from Example \ref{E:VR}. We have that the decomposition of $(\{H_0(V(X)(\epsilon))\}_{\epsilon \geq 0}\},\{H_0(\iota_{\epsilon, \epsilon'})\}_{\epsilon\leq \epsilon'})$ consists of the intervals $[0,0.5),[0,0.6),[0,1)$ and $[0,\infty)$. We can interpret each interval in the decomposition as describing the lifetime of one connected component, with the left enpoint representing the scale value at which it first appears, and the right endpoint represents the time at which it merges with another component: at scale $0$  the Vietoris--Rips simplicial complex consists only of  four $0$-simplices, and thus for $\epsilon=0$ we have four connected components. When we reach $\epsilon=0.5$ the components corresponding to $\{x_1\}$ and $\{x_3\}$ merge, and thus as a result we have that at this scale level there are only three connected components present. Similarly, we have that at scale  $0.6$ two further components merge, and similarly at scale value $1$, leaving only one connected component.

For $p=1$, the decomposition of $(\{H_1(V(X)(\epsilon))\}_{\epsilon \geq 0}\},\{H_1(\iota_{\epsilon, \epsilon'})\}_{\epsilon\leq \epsilon'})$ consists of a single interval $[1,1.1)$. Intuitively, this interval can be thought of as describing the lifetime of the square  formed by gluing together the four $1$-simplices $\{x_0,x_1\}, \{x_1,x_3\}, \{x_0,x_2\}, \{x_2,x_3\}$ as discussed in Example \ref{E:simpl hom}. 
The left endpoint of the interval is the lowest scale value at which the square appears in the collection of Vietoris--Rips complexes, while $1.1$ is the scale value at which it is filled in by the two  $2$-simplices $\{x_0,x_1,x_2\}$ and $\{x_1,x_2,x_3\}$.  

\end{example}

An alternative way of depicting the intervals in the decomposition in 
 Eq.~\ref{E:decomposition} is what is called a persistence diagram plot: 
this is a two-dimensional scatter plot in which each interval is represented by the point with coordinates given by its left and right endpoint. We note that in the barcode plot one needs to choose an order to stack the intervals, while in the persistence diagram plot one needs to choose a way to depict the multiplicity of the points.

\end{appendices}


%
%


\end{document}